\documentclass[sigconf,natbib=true]{acmart}

\AtBeginDocument{%
  \providecommand\BibTeX{{%
    \normalfont B\kern-0.5em{\scshape i\kern-0.25em b}\kern-0.8em\TeX}}}

\setcopyright{acmcopyright}
\copyrightyear{2018}
\acmYear{2018}
\acmDOI{XXXXXXX.XXXXXXX}

\acmConference[Conference acronym 'XX]{Make sure to enter the correct
  conference title from your rights confirmation emai}{June 03--05,
  2018}{Woodstock, NY}
\acmPrice{15.00}
\acmISBN{978-1-4503-XXXX-X/18/06}

\usepackage{subfigure}
\usepackage{float}
\usepackage{algorithm}
\usepackage{algpseudocode}
\usepackage{hyperref}
\usepackage{multirow}
\usepackage{adjustbox}

\usepackage[switch]{lineno}
\algnewcommand\algorithmicinput{\textbf{Input:}}
\algnewcommand\Input{\item[\algorithmicinput]}
\begin{document}
\title{Enhanced Bayesian Personalized Ranking for Robust Hard Negative Sampling in Recommender Systems}

\author{Kexin Shi}
\affiliation{%
  \institution{The Hong Kong University of Science\\ and Technology}
  \country{}}
\email{kshiaf@connect.ust.hk}
\author{Jing Zhang}
\affiliation{%
  \institution{The Hong Kong University of Science\\ and Technology}
  \country{}}
\email{jzhanggy@connect.ust.hk}
\author{Linjiajie Fang}
\affiliation{%
  \institution{The Hong Kong University of Science\\ and Technology}
  \country{}}
\email{lfangad@connect.ust.hk}
\author{Wenjia Wang}
\affiliation{%
  \institution{The Hong Kong University of Science\\ and Technology (Guangzhou)}
  \country{}}
\email{wenjiawang@hkust-gz.edu.cn}
\author{Bingyi Jing}
\affiliation{%
  \institution{Southern University of Science\\ and Technology}
  \country{}}
\email{jingby@sustech.edu.cn}


\begin{abstract}
In implicit collaborative filtering, hard negative mining techniques are developed to accelerate and enhance the recommendation model learning. However, the inadvertent selection of false negatives remains a major concern in hard negative sampling, as these false negatives can provide incorrect information and mislead the model learning. To date, only a small number of studies have been committed to solve the false negative problem, primarily focusing on designing sophisticated sampling algorithms to filter false negatives. In contrast, this paper shifts its focus to refining the loss function. We find that the original Bayesian Personalized Ranking (BPR), initially designed for uniform negative sampling, is inadequate in adapting to hard sampling scenarios. Hence, we introduce an enhanced Bayesian Personalized Ranking objective, named as Hard-BPR, which is specifically crafted for dynamic hard negative sampling to mitigate the influence of false negatives. This method is simple yet efficient for real-world deployment. Extensive experiments conducted on three real-world datasets demonstrate the effectiveness and robustness of our approach, along with the enhanced ability to distinguish false negatives.
\end{abstract}





\maketitle
\section{Introduction}
Implicit collaborative filtering (CF) in recommender systems is to leverage users’ implicit feedback, such as clicks, purchases or viewing history, to improve user experiences by offering personalized content recommendations~\cite{hu2008collaborative,pan2008one,liang2018variational,najafabadi2016systematic,lee2010collaborative}. Prior efforts in model architecture design, such as matrix factorization (MF)~\citep{DBLP:conf/kdd/Koren08,DBLP:conf/nips/SalakhutdinovM07,DBLP:journals/computer/KorenBV09,DBLP:journals/corr/DziugaiteR15,lee2000algorithms,mnih2007probabilistic} and graph neural network (GNN) based methods~\citep{DBLP:conf/kdd/Wang00LC19, DBLP:journals/corr/BergKW17,DBLP:conf/sigir/Wang0WFC19, DBLP:conf/sigir/0001DWLZ020,DBLP:conf/kdd/HuangDDYFW021}, have effectively improved personalized recommendation accuracy. Nevertheless, a gap remains in the data domain, specifically in addressing the challenge presented by the lack of explicit negative feedback during recommendation model training. To address this, Bayesian Personalized Ranking (BPR)~\cite{rendle2012bpr} formulates the training objective as a pairwise ranking task, which aims to maximize the probability of user preferring observed positive items than randomly sampled negative items. In BPR, the logistic sigmoid function is used to model the individual preference probability, which almost remains a fixed paradigm in subsequent studies. However, random sampling to generate negative samples presents low convergence due to gradients vanishing under the BPR loss. Subsequently, hard negative sampling methods are developed to adaptively choose challenging items as negatives, compelling the model to achieve faster and more effective learning~\cite{zhao2023augmented,fan2023batch,zhang2013optimizing,rendle2014improving,huang2021mixgcf,park2019adversarial,shi2022soft,chen2023revisiting,wu2023dimension}. Various hard mining techniques, such as dynamic negative sampling~\cite{zhang2013optimizing}, Generative Adversarial Network (GAN) based sampling~\cite{wang2017irgan,park2019adversarial} and importance sampling based hard mining~\cite{chen2022learning,lian2020personalized}, emerge. Among hard sampling methods focusing on sampling progress, many adopt the BPR loss for model optimization. Conversely, for hard sampling methods that improve loss function, it is prevalent to apply importance sampling to the BPR loss but keep the core BPR unchanged. The question is: \textit{Does the original BPR effectively accommodate hard sampling scenarios?}

In practice, the application of hard sampling in optimizing the recommendation model under the original BPR objective has been reported to potentially result in overfitting~\cite{shi2022soft,zhu2022gain}. This is because false negatives, inadvertently selected during hard negative sampling, contribute significant gradients to model updating and potentially mislead the learning direction. To address the false negative problem and improve model’s robustness, SRNS~\cite{ding2020simplify} introduces a variance-based sampling strategy to identify real hard negatives and GDNS~\cite{zhu2022gain} considers the expectation gain between two training iterations as an indicator. Both methods concentrate on crafting sophisticated sampling processes to mitigate the risk of sampling false negatives. However, achieving this task solely through a single metric, such as variance or expectation gain, is challenging.

According to the theory that the effects of a specific sampling strategy on model learning can be equivalently achieved by adjusting the loss function~\cite{fujimoto2020equivalence}, our focus shifts from designing intricate sampling process for filtering false negatives towards crafting the loss function. We introduce an enhanced BPR scheme, named Hard-BPR, to mitigate the influence of false negatives in hard sampling. Compared to BPR, only the function to estimate the individual preference probability is modified in Hard-BPR, where three coefficients are added to lower the gradient magnitudes of excessive hard samples when applying stochastic gradient descent. The framework is straightforward, employing the hard negative sampling method DNS~\cite{zhang2013optimizing} to sample hard negatives and utilizing Hard-BPR as the optimization objective. Since the negative sampling process, the primary time-consuming element, remains DNS, our algorithm achieves superior time efficiency compared to other hard sampling approaches. Moreover, a recent study discloses that employing DNS under BPR is an exact estimator of One-way Partial AUC (OPAUC)~\cite{shi2023theories}. We show that equipping DNS under the proposed Hard-BPR also serves as an exact OPAUC estimator. We conduct experiments on three real-world datasets, evaluating the effects of Hard-BPR comprehensively. Our results illustrate its enhanced ability to distinguish false negatives from real hard negatives, along with its effectiveness and efficiency in recommendation model training. In addition, a parameter study on three critical coefficients in Hard-BPR reveals that only two coefficients require fine-tunning, providing valuable guidance for algorithm deployment. 

In summary, the contributions of this paper are:
\begin{itemize}
    \item We are, to the best of our knowledge, the first to address the false negative problem in implicit CF from the perspective of BPR redesign.
    \item We introduce an optimization criterion, Hard-BPR, an enhanced Bayesian Personalized Ranking for dynamic hard negative sampling (DNS) in recommender systems. 
    \item Our experiments empirically show the proposed method’s robustness, efficiency and effectiveness. Moreover, parameter study offers valuable insights for its implementation. 
\end{itemize}

\section{Preliminaries}
In this section, we present the necessary background about the derivation of BPR in implicit collaborative filtering and dynamic hard negative sampling (DNS).
\subsection{Bayesian Personalized Ranking (BPR)}
\label{sec:BPR}
In implicit collaborative filtering, only the implicit feedback $\mathcal{S} \subseteq \mathcal{U} \times \mathcal{I}$ is observed (such as clicks, views, or purchases), where $\mathcal{U}$ denotes the user set and $\mathcal{I}$ the item set. The user-item interaction pairs in implicit feedback $(u,i) \in \mathcal{S}$ are positive observations because they generally indicate users’ preferences.  The remaining unobserved user-item combinations $(\mathcal{U} \times \mathcal{I}) \setminus \mathcal{S}$ consist of actual negative feedback (indicating dislikes) and the potential positive feedback (implying future purchase interests). 

For each user $u$, we define $\mathcal{I}^+(u):=\left\{i : (i,u)\in \mathcal{S}\right\}$. The set $\mathcal{I}^+(u)$ contains all items that interact with user $u$. BPR assumes that user $u$ prefers item $i\in\mathcal{I}^+(u)$ over $j\in \mathcal{I}\setminus\mathcal{I}^+(u) $ (symbolized as $i >_{u} j$). The objective of BPR is to learn a scoring function $f(\cdot|\Theta): \mathcal{I}\times\mathcal{U} \to \mathbb{R}$ to capture users’ above pairwise preferences, where $f(i|u,\Theta)$ is the score of item $i$ given by user $u$ and $\Theta$ is the model parameter. We denote $\mathcal{D}_{\mathcal{S}}=\left\{(u,i,j) \mid i \in \mathcal{I}^{+}(u), j \in \mathcal{I} \setminus \mathcal{I}^{+}(u)\right\}$. The likelihood function of observing all user pairwise preferences in $\mathcal{D}_{\mathcal{S}}$ given the model parameters $\Theta$ is as follows:

\begin{equation}
\prod_{(u, i, j) \in D_S} P(i >_u j | \Theta),
\end{equation}
where $P(i >_u j | \Theta)$ is the probability that user $u$ prefers item $i$ over item $j$.
The individual preference probability $P(i >_u j | \Theta)$ is modeled as:

\begin{equation}
P(i >_u j | \Theta) = \sigma(\hat{x}_{uij}),
\end{equation}
where $\hat{x}_{uij}$ represents the estimated preference difference between item $i$ and item $j$, which can be calculated as $f(i|u,\Theta) - f(j|u,\Theta)$, and $\sigma$ is the logistic sigmoid function:
\begin{equation}
\sigma(x):=\frac{1}{1+e^{-x}}. 
\end{equation}
The logistic sigmoid function $\sigma$ is used to transform the estimated preference difference between two items into a probability between 0 and 1. This non-linear transformation can help capture the complex nature of user preferences in recommender systems.

Then, the BPR loss is derived as the negative log-likelihood of the pairwise preferences in $\mathcal{D}_{\mathcal{S}}$:
\begin{equation}
\text{L}_\text{BPR} = -\sum_{(u, i, j) \in D_S} \ln \sigma(\hat{x}_{uij}),
\end{equation}
In the context of optimizing the Bayesian Personalized Ranking (BPR) objective, Stochastic Gradient Descent (SGD) is employed, wherein triplets $(u,i,j)$ are uniformly drawn from $\mathcal{D}_\mathcal{S}$ to facilitate the updating of model parameters. Specifically, for each positive user-item interaction $(u,i)$, a negative item $j$ is randomly selected from the set $\mathcal{I} \setminus \mathcal{I}^+(u)$. The gradient of the BPR loss with respect to a model parameter $\theta \in \Theta$ is as follows:
\begin{equation}
\frac{\partial\text{L}_\text{BPR}}{\partial\theta} = -\sum_{(u, i, j) \in D_S}\left(1-\sigma\left(\hat{x}_{uij}\right)\right) \frac{\partial\left(\hat{x}_{uij}\right)}{\partial \theta},
\end{equation}
The impact of each triplet $(u,i,j)$ on the model's update is quantified by the gradient magnitude $\Delta_{\sigma}(\hat{x}_{uij})$, expressed as:
\begin{equation}
\Delta_{\sigma}(\hat{x}_{uij}):=(1-\sigma(\hat{x}_{uij}))=\left(1-P(i >_u j | \Theta)\right)
\end{equation} 

\citet{rendle2014improving} suggests that uniform sampling strategy results in slow convergence of the model because most of samples yield small gradient magnitudes. Specifically, the scoring function $f(\cdot|\Theta)$ generally gives a larger score to positive pair $(u,i)$ than pair $(u,j)$ and the pairwise preference probability $P(i >_u j | \Theta)$ is close to 1.

\subsection{Dynamic Hard Negative Sampling}
To accelerate the model's convergence and enhance its ability to differentiate between positive and negative items, various non-uniform sampling methods have been proposed. Among these, hard negative samplers demonstrate notable efficiency. Hard negative sampling is to choose negative instances that are challenging for model to distinguish from positive examples, thereby enhancing the model’s capacity to learn fine-grained features and improving its discriminative ability. 

Dynamic Negative Sampling (DNS)~\cite{zhang2013optimizing}, supported by various studies~\cite{huang2021mixgcf,ding2020simplify,shi2022soft}, stands as an advanced hard negative sampling method in recommender systems. \citet{shi2023theories} has proven that the BPR loss equipped with DNS serves as an exact estimator of OPAUC. DNS involves a two-step process. This two-step process begins by selecting $H (\geq 1)$ random items from the set $\mathcal{I} \setminus \mathcal{I}^+(u)$ for each positive user-item pair $(u,i)$, creating a negative candidate pool $\mathcal{P} = \left\{j_1, j_2, \ldots, j_H\right\}$. Then, the item from $\mathcal{P}$ that is assigned the highest score by the scoring function $f(\cdot|u,\Theta)$ is selected as the hard negative sample. This is represented as:
\begin{equation}
j = \mathop{\arg\max}_{k \in \mathcal{P}}f(\cdot|u,\Theta),
\label{step:DNS}
\end{equation}
The selection of a hard negative item results in a relatively large gradient magnitude, $\Delta_{\sigma}(\hat{x}_{uij})$, for the triplet $(u,i,j)$ since the estimated preference difference between the positive item $i$ and the hard negative item $j$ is small. Consequently, this leads to a more significant contribution to the model's updating process. By efficiently choosing hard negative samples, DNS accelerates training processes and improves prediction accuracy.

\section{method: Hard-BPR}
In this section, we introduce the false negative problem for hard negative sampling in implicit CF. Subsequently, we propose an enhanced Bayesian Personalized Ranking, termed as Hard-BPR, to address the challenges posed by false negatives.

\subsection{Challenges of False Negatives}
\label{sec:false_negative}

A critical challenge in hard negative sampling within recommender systems is the risk of inadvertently including false negatives, which may induce overfitting, potentially compromising the model’s generalization ability. False negatives in recommender systems are items that the user has not interacted with, but the user would have liked or found interesting. As the hardness of the negatives sampled increases, the probability of encountering false negatives correspondingly increases~\cite{shi2022soft, bekker2020learning,cai2022hard}. To be specific, let \( H(j) \) denote the hardness of a negative item \( j \), and \( P_{\text{FN}}(j) \) be the probability that item \( j \) is a false negative. As \( H(j) \) increases, \( P_{\text{FN}}(j) \) also rises. This relationship poses a significant limitation to hard negative sampling methods, as it is challenging to distinguish between real hard negatives and false negatives.

The corruption of false negatives not only reduces the accuracy of personalized recommendations but also exacerbates overfitting during the model training~\cite{shi2022soft,zhu2022gain}. Avoiding false negatives in hard negative sampling is crucial, yet only a limited number of studies focus on developing robust hard sampling methods.

\subsection{Hard-BPR}

\begin{figure}[t]
\centering
\subfigure[]{
\label{fig.g(x)}
\includegraphics[width=0.45\columnwidth]{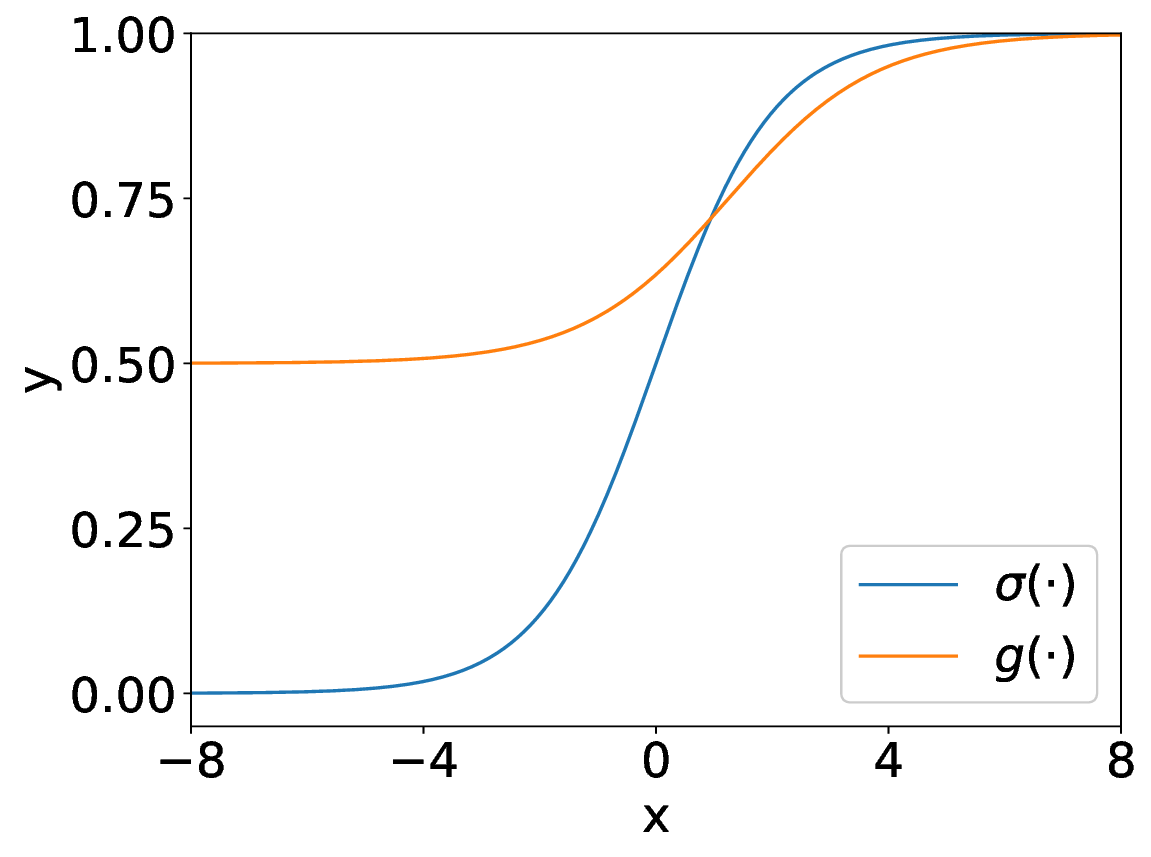}}
\subfigure[]{
\label{fig.Delta(x)}
\includegraphics[width=0.45\columnwidth]{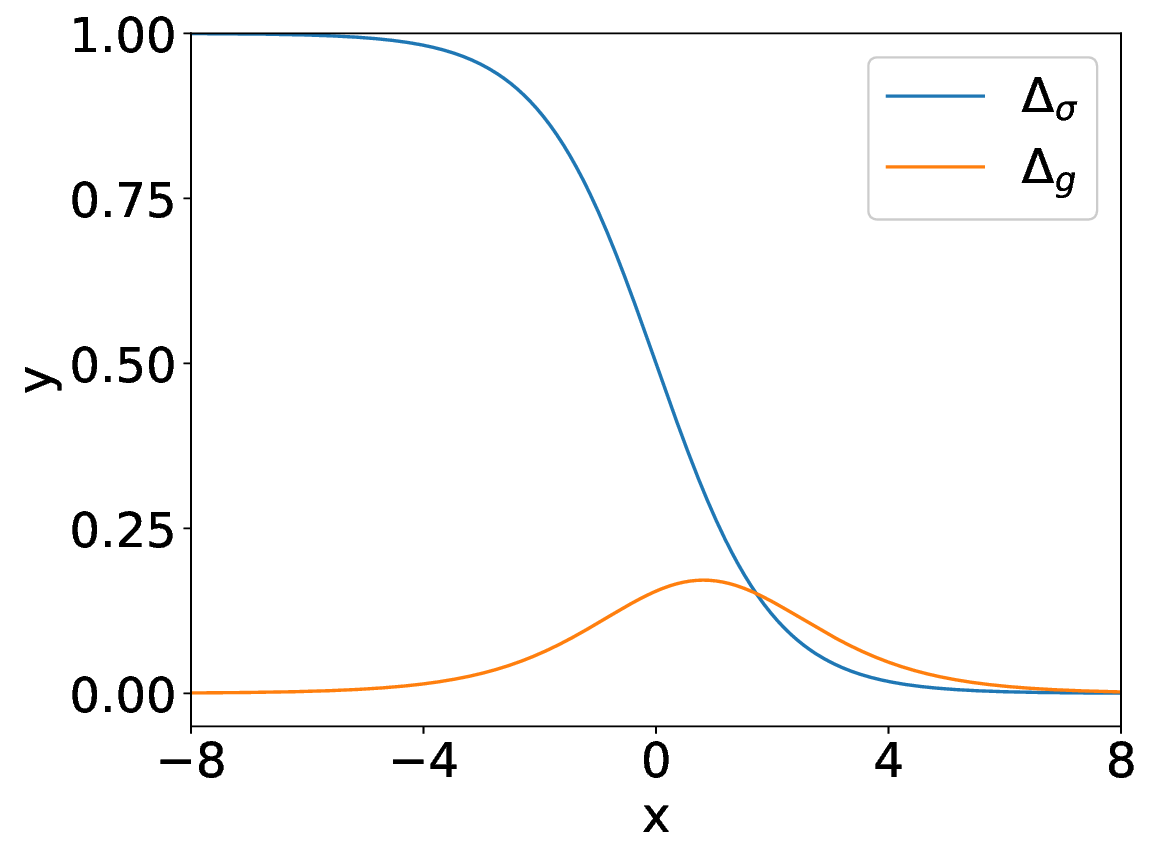}}
\caption{(a) The function curves of $\sigma(\cdot)$ and $g(\cdot)$. (b) The function curves of $\Delta_\sigma(\cdot)$ and $\Delta_g(\cdot)$.}
\end{figure}
\textbf{The Equivalence Theory.}
The foundational principle guiding our methodology is derived from the equivalence theory between loss functions and sampling strategies. This theory indicates that the effects induced by a specific sampling strategy on model learning can be equivalently achieved through modifying the loss function~\cite{fujimoto2020equivalence}. 

In implicit CF task, previous works focus on designing complex and sophisticated sampling algorithms to avoid false negatives in hard negative sampling. The equivalence theory suggests that an alternate loss function, when optimized with a simple sampling method (\emph{e.g.}, DNS), can yield an expected gradient that is similar to the one obtained with a complex false-negative-exclusive hard sampling approach. The equivalence theory can be mathematically expressed in terms of the expected gradient of the loss function, as follows:
\begin{equation}
E_{(u,i,j)\sim f_{\text{complex}}}[\nabla L_{\text{BPR }}(u,i,j)] = E_{(u,i,j)\sim f_{\text{DNS }}}[\nabla L_{\text{mod}}(u,i,j)].
\end{equation}
Here, $E_{(u,i,j)\sim f_{\text{complex}}}[\nabla L_{\text{BPR }}(u,i,j)]$ denotes the expected gradient of BPR loss under the sophisticatedly deigned false-negative-exclusive hard sampling, and $E_{(u,i,j)\sim f_{\text{DNS }}}[\nabla L_{\text{mod}}(u,i,j)]$ represents the expected gradient of a modified loss function under DNS.

This theory suggests a new perspective on addressing the challenges of hard negative sampling in implicit CF. Instead of creating complex sampling algorithms to discern and exclude false negatives, the equivalence theory encourages a shift towards revising the loss function.
\\\\
\textbf{Limitations of BPR in Hard Negative Sampling.}
As mentioned in Section~\ref{sec:BPR}, BPR employs the logistic sigmoid function $\sigma (\cdot)$ to calculate the individual preference probability $P(i>_u j|\Theta)$. When employing BPR, $P(i >_u j | \Theta)$ approaches 0 if $\hat{x}_{uij}$ is small (\emph{e.g.}, $\leq-4$). In this situation, the gradient magnitude $\Delta_{\sigma}(\hat{x}_{uij})$ is near 1, and thus the triplet $(u, i, j)$ can provide useful information for model optimization via SGD. Conversely, a large $\hat{x}_{uij}$ results in a negligible $\Delta_{\sigma}(\hat{x}_{uij})$, rendering the triplet $(u, i, j)$ useless in parameter updating. 

\citet{rendle2014improving} states that $\hat{x}_{uij}$ is generally a large value when item $j$ is sampled uniformly from $\mathcal{I}\setminus\mathcal{I}^+(u)$, and only a few $(u, i, j)$ triplets may yield small $\hat{x}_{uij}$s. In this case, the BPR loss is suitable, since it can assign large gradient magnitudes $\Delta_{\sigma}(\hat{x}_{uij})$ to these few triplets, thereby expediting the model convergence.

However, when adopting hard negative sampling strategies such as DNS for negative selection, the $\hat{x}_{uij}$ of triplet $(u, i, j)$ is typically small, and items $j$ yielding smaller values  are more likely to be false negatives, as discussed in Section~\ref{sec:false_negative}. In this case, the logistic sigmoid function in BPR loss is inappropriate because it easily generates a large $\Delta_{\sigma}(\hat{x}_{uij})$ for false negatives and misleading the parameter updating. 
\\\\
\textbf{The Proposed Method.}
\begin{table*}[htp]
\caption{Comparison of the proposed method and baselines, where $\mathcal{S}$ is the training set containing user-item interactions, $\mathcal{U}$ is the user set, $\mathcal{I}$ is the item set, $H_1$ and $H_2$ denote the negative candidate pool size, $T$ is the time to compute user-item score and $T_Q$ is the sampling time for a proposal distribution in AdaSIR.}
\begin{tabular}{cccc}
\hline & Learning Objective & Time Complexity & Robustness to FN \\
\hline RNS & BPR & $\mathrm{O}(|\mathcal{S}| T)$ & $\times$ \\
\hline AdvIR & adversarial loss & $\mathrm{O}\left(|\mathcal{S}| H_1 T\right)$ & $\times$ \\
\hline IRGAN & adversarial loss & $\mathrm{O}(|\mathcal{S}||\mathcal{I}| T)$ & $\times$ \\
\hline DNS & BPR & $\mathrm{O}\left(|\mathcal{S}| H_1 T\right)$ & $\times$ \\
\hline SRNS & BPR & $\mathrm{O}\left(|\mathcal{S}|\left(H_1+H_2\right) T\right)$ & $\sqrt{ }$ \\
\hline GDNS & group-wise ranking loss & $\mathrm{O}\left(|\mathcal{U}|\left(2 H_1 T+H_1^2\right)\right)$ & $\sqrt{ }$ \\
\hline MixGCF & BPR & $\mathrm{O}\left(|\mathcal{S}| H_1 T\right)$ & $\times$ \\
\hline AdaSIR & importance sampling BPR & $\mathrm{O}\left(|\mathcal{U}| H_1\left(T_Q+T\right)+|\mathcal{S}| H_2 T\right)$ & $\times$ \\
\hline Ours & Hard-BPR & $\mathrm{O}\left(|\mathcal{S}| H_1 T\right)$ & $\sqrt{ }$\\
\hline
\end{tabular}
\label{tab:time_complexity}
\end{table*}
The proposed framework is shown in Algorithm~\ref{alg:hard_bpr}. To alleviate the negative effects of false negatives on model learning, we propose Hard-BPR, an enhanced Bayesian Personalized Ranking. This is achieved by substituting $\sigma (\cdot)$ with $g(\cdot)$ for the estimation of individual preference probability. The function $g(\cdot)$ is a transformation of $\sigma(\cdot)$ and its mathematical expression is as below:
\begin{equation}
g(x) = \frac{1}{1+a}[\sigma(cx+b)+a],
\end{equation}
where $a, b\text{ and } c$ are constant coefficients with $a\geq 0$ and $c>0$. The Hard-BPR loss is defined as:
\begin{equation}
\text{L}_\text{Hard-BPR} = -\sum_{\substack{(u,i)\in \mathcal{S}\\j\sim f_{\mathrm{DNS}}}} \ln g(\hat{x}_{uij}).
\end{equation}
Here $j\sim f_{\mathrm{DNS}}$ represents that item $j$ is sampled as a negative by hard negative sampling strategy DNS. The Hard-BPR loss proposed in this study is an exact estimator of OPAUC when equipped with DNS, because $-\ln g(\cdot)$ ($a\geq 0$, $c>0$) is a convex, differentiable and monotonically decreasing function, satisfying the sufficient condition to be consistent for OPAUC maximization according to works~\cite{gao2012consistency,shi2023theories}. OPAUC denotes the partial area under the ROC curve, wherein the false positive rate (FPR) is constrained within the range $[0, \beta]$ ($0<\beta\leq 1$). According to \citet{shi2023theories}, optimizing OPAUC can lead to improved Top-$K$ recommendation performance when the parameter $\beta$ tuned appropriately. This optimization strategy allows for emphasizing the ranking of top-ranked items instead of the whole ranking list.

For the purpose of model optimization under SGD, the gradient of Hard-BPR loss is calculated as:
\begin{equation}
\begin{aligned}
\frac{\partial\text{L}_\text{Hard-BPR}}{\partial\theta} &= -\sum_{\substack{(u,i)\in \mathcal{S}\\j\sim f_{\mathrm{DNS}}}}\Delta_{g}(\hat{x}_{uij})\frac{\partial\left(\hat{x}_{uij}\right)}{\partial \theta}, \\
\text{with  } \Delta_{g}(\hat{x}_{uij}) &= \frac{c\sigma(c\hat{x}_{uij}+b)(1-\sigma(c\hat{x}_{uij}+b))}{\sigma(c\hat{x}_{uij}+b)+a}.
\end{aligned}
\end{equation}
\\\\
\textbf{Properties.} The coefficient $a$ in $g(\cdot)$ is the most critical coefficient, which is essential for converting the magnitude gradient function from a monotonically decreasing pattern to unimodal symmetry, as illustrated in Figure~\ref{fig.g(x)}. $\Delta_{g}(\cdot)$ is bell-shaped under the condition $a>0$. Additionally, the coefficients $b \text{ and } c$ are for $\Delta_{g}(\cdot)$ translation and stretching. The function $\Delta_{g}(\cdot)$ derived from our new individual preference probability estimator $g(\cdot)$ has following properties:
\begin{itemize}
    \item {The limit of $\Delta_g(x)$ as $x$ approaches either positive or negative infinity is equal to 0.}
    \item {$\Delta_g(x)$ is unimodal and symmetric with respect to the maximum point located at $x$ (see Appendix for more details).}
    \item {$\Delta_g(x)$’s maximum point $x_{\text{max}}$ and maximum value $\Delta_{g,\text{max}}$ are calculated as:}
\end{itemize}
\begin{equation}
x_{\text{max}} = \frac{-b+\ln(\frac{\sqrt{a}}{\sqrt{1+a}})}{c},
\label{eq:x_max}
\end{equation}

\begin{equation}
\Delta_{g,\text{max}} = \frac{\sqrt{1 + a} \, c}{2 \sqrt{a} + 2 a^{3/2} + \sqrt{1 + a} + 2 a \sqrt{1 + a}}.
\end{equation}
\\\\
\textbf{Mechanism Analysis.} We analyze the resilience of Hard-BPR to false negatives in hard negative sampling. We illustrate this by comparing the functions $\sigma(\cdot)$ and $g(\cdot)$ in Figure~\ref{fig.g(x)}, along with their respective gradient magnitudes $\Delta_{\sigma}(\cdot)$ and $\Delta_{g}(\cdot)$ in Figure~\ref{fig.Delta(x)}. For $g(\cdot)$, we set parameters $a, b\text{ and } c$ to $1,-1\text{ and }0.8$.

The foundational principle of designing $g(\cdot)$ is to moderate the weight assigned to excessively hard negatives in model learning. Specifically, if a negative item $j$ chosen by DNS for the user-item pair $(u,i)$ is excessively hard (characterized by a very small $\hat{x}_{uij}$), we avoid modeling $P(i >_u j | \Theta)$ as nearly zero, as is done with $\sigma(\cdot)$. Instead, using $g(\cdot)$, we assign a small positive value to $P(i >_u j | \Theta)$. In this way, the gradient magnitude $\Delta_{g}(u, i, j)$ is reduced to a relatively small value, thereby decreasing its influence on model updating in cases where the hard negative instance is a false negative.

The Hard-BPR loss allows more flexibility than the BPR loss. The gradient magnitude curve derived from $g(\cdot)$ is single-peaked, by adjusting the constants $a, b\text{ and } c$ to change the position of peak value, the model can focus on learning from a specific hardness level of selected negatives. Note that setting $a, b \text{ and } c$ to $(0,0,1)$ reduces the Hard-BPR loss to the original BPR loss.

\begin{algorithm}[h]
\caption{Optimizing models with Hard-BPR}
\label{alg:hard_bpr}
\begin{algorithmic}[1]
\Input Training set $\mathcal{S}=\left\{(u,i)\right\}$, the scoring function $f(\cdot|\Theta)$ with learnable parameter $\Theta$, the size of negative candidate pool $H$, constant coefficients $a, b$ and $c$ in the Hard-BPR loss.
\For{$t=1,2,\cdots,T$}
\State Sample a mini-batch $\mathcal{S}_{batch}\in \mathcal{S}$
\For{each $(u,i) \in \mathcal{S}_{batch}$}
\State Randomly sample $H$ items from $\mathcal{I} \setminus \mathcal{I}_u$ to form the
\State negative candidate pool $\mathcal{P}$ for user $u$;
\State Choose the hard negative item $j$ by DNS;
\State Update $\text{L}_{\text{Hard-BPR}}$.
\EndFor
\State Optimize $\Theta$ based on $\text{L}_{\text{Hard-BPR}}$ with SGD algorithm.
\EndFor
\end{algorithmic}
\end{algorithm}

\subsection{Complexity analysis}

\textbf{Time Complexity.} We conduct a time complexity comparison between the proposed algorithm and a set of state-of-the-art negative sampling approaches, as outlined in Table~\ref{tab:time_complexity}. The proposed method requires lower time complexity among the considered baselines. For each training epoch, the proposed method takes a computational cost of $O(|\mathcal{S}| H T)$, where $|\mathcal{S}|$ denotes the number of positive user-item observations in set $\mathcal{S}$, $H$ represents the number of negative candidates selected for each positive user-item pair, and $T$ denotes the runtime for user-item score computation. The time complexity of the proposed method is the same with DNS, since only the function for estimating pairwise preferences has been modified. Both DNS and our method stand out as two of the most efficient algorithms in hard negative mining. In comparison, other methods such as IRGAN, SRNS, GDNS, and AdaSIR, which design specific sampling procedures, demand relatively higher computational resources.
\\
\textbf{Space Complexity.} In addition to model parameters, the space complexity of the proposed method comes from storing negative candidates for each positive user-item pair, incurring a space complexity of $O(|\mathcal{S}| H)$.

\section{Experiments}
In this section, we evaluate the performance of our approach on three real-world datasets, with MF and LightGCN as the base recommendation models. Additionally, we study the proposed method’s ability to discern false negatives and investigate its behavior under different parameter settings. 

\begin{table}
\caption{Statistics of datasets.}
\begin{adjustbox}{width=\columnwidth}
\begin{tabular}{crrrrrr}
\hline Dataset & \#User & \multicolumn{1}{c}{ \#Item } & \multicolumn{1}{c}{ \#Train } & \multicolumn{1}{c}{ \#Val } & \multicolumn{1}{c}{ \#Test } & Density \\
\hline Taobao & 22,976 & 29,149 & $351,444$ & $43,930$ & $43,931$ & $0.00066$\\
Tmall & 10,000 & 14,965 & $369,520$ & $46,189$ & $46,190$ & $0.00309$\\
Gowalla & 29,858 & 40,981 & $821,896$ & $102,737$ & $102,737$ & $0.00084$\\
\hline
\end{tabular}
\end{adjustbox}
\label{tab:dataset}
\end{table}

\subsection{Experimental Settings}
\textbf{Datasets.} We assess our method on three real-world datasets: Taobao\footnote{\url{https://tianchi.aliyun.com/dataset/dataDetail?dataId=649}}, Tmall\footnote{\url{https://tianchi.aliyun.com/dataset/dataDetail?dataId=121045}} and Gowalla. Taobao and Tmall, sourced from e-commerce platforms, capture diverse user behaviors such as clicks, cart additions, and purchases. To address data sparsity, varied interaction behaviors are considered positive labels in the prediction task. The 10-core setting is applied by randomly selecting users with a minimum of 10 interaction records. Gowalla consists of check-in records with user locations and we directly utilize the pre-processed dataset in work~\cite{DBLP:conf/sigir/Wang0WFC19} for model performance evaluation. For each dataset, we construct training, validation, and test sets by sorting interactions based on timestamps and retaining the latest 10\% of records for testing, with the remaining interactions divided into an 80/10 ratio for training and validation (see details in Table~\ref{tab:dataset}). 
\\\\
\textbf{Metrics.} The main aim of personalized recommendation systems is to provide a ranked list containing the Top-$K$ items with the highest scores for each user. This study uses two widely-adopted metrics, Recall@$K$ and NDCG@$K$, to evaluate the model's ability to understand users' preferences. The work by \citet{shi2023theories} reports the connection between $K$ in Top-$K$ metrics and the size of the negative candidate pool ($H$) in DNS, which suggests that the lower the $K$, the larger the $H$ when the training curve reaches its maximum values. To control memory usage related to storing negative candidates, we set $K$ to 50 as done in previous works~\citep{chen2022learning,lian2020personalized,shi2023theories}.
\\\\
\textbf{Baselines.} A variety of negative samplers in recommender systems are chosen as baselines to assess the effectiveness of the proposed method. The selected baselines are detailed as follows:
\begin{itemize}
\item {\textbf{RNS}~\cite{rendle2012bpr}} selects negative instances by a uniform distribution, which is widely employed in implicit CF studies.
\item {\textbf{AdvIR}~\cite{park2019adversarial}} incorporates Generative Adversarial Nets to produce challenging negative instances, and utilizes virtual adversarial training to enhance model performance. 
\item {\textbf{IRGAN}~\cite{wang2017irgan}} is also a GAN-based hard negative sampler and optimizes model parameters through a min-max game between the generative retrieval network and the discriminative network. 
\item {\textbf{DNS}~\cite{zhang2013optimizing}} first randomly samples a set of negative candidates for each positive instance, then chooses the highest-scored candidate as the hard negative sample.
\item {\textbf{SRNS}~\cite{ding2020simplify}} considers the variance of item scores during training to avoid false negatives in the hard sampling procedure.
\item {\textbf{GDNS}~\cite{zhu2022gain}} employs a gain-ware sampler to reduce the probability of introducing false negatives during hard negative sampling.
\item {\textbf{MixGCF}~\cite{huang2021mixgcf}} introduces hop-mixing and positive mixing techniques to synthesized hard negative samples for GNN-based recommender models.  
\item {\textbf{AdaSIR}~\cite{chen2022learning}} reuses informative negatives and adopts importance resampling to approximate the softmax distribution for effectively selecting hard negatives. AdaSIR(U) and AdaSIR(P) represent sampling negative candidate using a uniform distribution and a popularity distribution respectively. \\
\end{itemize}

\begin{table*}[htp]
  \centering
  \caption{Performance comparison. The best results are presented in bold and the second best results are underlined. The relative improvements of the best results compared to the second best results are listed in the last row.}
    \begin{tabular}{r|cc|cc|cc|cc|cc|cc}
    \toprule
          & \multicolumn{6}{c|}{\textbf{MF}}              & \multicolumn{6}{c}{\textbf{LightGCN}} \\
\cmidrule{2-13}          & \multicolumn{2}{c|}{Taobao} & \multicolumn{2}{c|}{Tmall} & \multicolumn{2}{c|}{Gowalla} & \multicolumn{2}{c|}{Taobao} & \multicolumn{2}{c|}{Tmall} & \multicolumn{2}{c}{Gowalla} \\
          & \multicolumn{1}{c}{Recall} & \multicolumn{1}{c|}{NDCG} & \multicolumn{1}{c}{Recall} & \multicolumn{1}{c|}{NDCG} & \multicolumn{1}{c}{Recall} & \multicolumn{1}{c|}{NDCG} & \multicolumn{1}{c}{Recall} & \multicolumn{1}{c|}{NDCG} & \multicolumn{1}{c}{Recall} & \multicolumn{1}{c|}{NDCG} & \multicolumn{1}{c}{Recall} & \multicolumn{1}{c}{NDCG} \\
    \midrule
    \multicolumn{1}{l|}{RNS} & 0.0751 & 0.0268 & 0.0692 & 0.0333 & 0.2230 & 0.1418 & 0.1041 & 0.0378 & 0.0895 & 0.0476 & 0.2712 & 0.1804 \\
    \multicolumn{1}{l|}{AdvIR} & 0.0911 & 0.0327 & 0.0792 & 0.0426 & 0.2603 & 0.1620 & 0.1052 & 0.0382 & 0.0917 & 0.0487 & 0.2710 & 0.1801 \\
    \multicolumn{1}{l|}{IRGAN} & 0.0859 & 0.0314 & 0.0771 & 0.0415 & 0.2421 & 0.1508 & 0.1049 & 0.0380 & 0.0904 & 0.0480 & 0.2799 & 0.1847 \\
    \multicolumn{1}{l|}{DNS} & \underline{0.0999} & \underline{0.0360} & 0.0688 & 0.0320 & 0.2490 & 0.1582 & 0.1107 & \underline{0.0410} & 0.0970 & 0.0504 & \underline{0.2848} & \underline{0.1878} \\
    \multicolumn{1}{l|}{SRNS} & 0.0968 & 0.0349 & 0.0801 & 0.0433 & 0.2588 & 0.1620 & 0.1086 & 0.0392 & 0.0967 & 0.0509 & 0.2819 & 0.1866 \\
    \multicolumn{1}{l|}{GDNS} & 0.0970 & 0.0352 & \underline{0.0816} & \underline{0.0440} & 0.2611 & 0.1638 & 0.1089 & 0.0393 & 0.0975 & 0.0516 & 0.2835 & 0.1865 \\
    \multicolumn{1}{l|}{MixGCF} &   -    &      - &    -   &    -   &   -    &    -   & \underline{0.1110} & 0.0403 & 0.0963 & 0.0508 & 0.2840 & 0.1872 \\
    \multicolumn{1}{l|}{AdaSIR(U)} & 0.0958 & 0.0336 & 0.0800 & 0.0429 & \underline{0.2711} & \underline{0.1712} & 0.1100 & 0.0401 & 0.1009 & 0.0530 & 0.2816 & 0.1857 \\
    \multicolumn{1}{l|}{AdaSIR(P)} & 0.0955 & 0.0333 & 0.0810 & 0.0438 & 0.2702 & 0.1700 & 0.1080 & 0.0389 & \underline{0.1016} & \underline{0.0538} & 0.2808 & 0.1853 \\
    \midrule
    \multicolumn{1}{l|}{Ours} & \textbf{0.1055} & \textbf{0.0386} & \textbf{0.0876} & \textbf{0.0480} & \textbf{0.2825} & \textbf{0.1813} & \textbf{0.1150} & \textbf{0.0419} & \textbf{0.1054} & \textbf{0.0556} & \textbf{0.2976} & \textbf{0.1947} \\
          & 5.61\% & 7.22\% & 7.35\% & 9.09\% & 4.21\% & 5.90\% & 3.60\% & 2.20\% & 3.74\% & 3.35\% & 4.49\% & 3.67\% \\
    \bottomrule
    \end{tabular}%
  \label{tab:comparison_mf_lgn}%
\end{table*}%
\noindent\textbf{Implementation Details.} We perform experiments using PyTorch on a single Linux server featuring an AMD EPYC 7543 processor, 128GB RAM, and an NVIDIA GeForce RTX 4090 GPU. Adam~\citep{kingma2014adam} is employed as the optimizer, and a fixed batch size of 2048 is utilized across all experiments. Without loss of generality, our underlying recommendation models include matrix factorization~\cite{koren2008factorization} (MF) and LightGCN~\citep{he2020lightgcn}, where LightGCN is a state-of-the-art GNN-based framework. For the proposed method, we first conduct grid search to determine optimal hyperparameters for DNS. Subsequently, Bayesian optimization is employed to explore coefficients $a, b, c$ in $\mathrm{L}_{\mathrm{Hard-BPR}}$. User and item embedding dimensions are tuned from the set $\left\{8,16,32,64\right\}$, while the negative candidate pool size for our method is searched in $\left\{8,16,32,64\right\}$. $L_2$ regularization is adjusted within $\{0,0.1,0.01,0.001,0.0001\}$, and the learning rate is explored within $\{0.0005,0.001,0.005,0.01\}$. Coefficients $a, b\text{ and } c$ in $\text{L}_\text{Hard-BPR}$ are uniformly searched in the ranges $[0, 10]$, $[-10, 10]$, and $(0, 5]$, respectively. Hyperparameters are tuned to optimize Recall@50 on the validation set, and final results are reported on the test set. To ensure a fair comparison, all baselines are finely tunned based on the best Recall@50 achieved on the validation set. Additionally, two GAN-based samplers are trained from a uniformly pretrained model.

\subsection{Performance comparison}
\begin{figure*}[t]
\centering
\subfigure[MF, Taobao]{
\includegraphics[width=0.55\columnwidth]{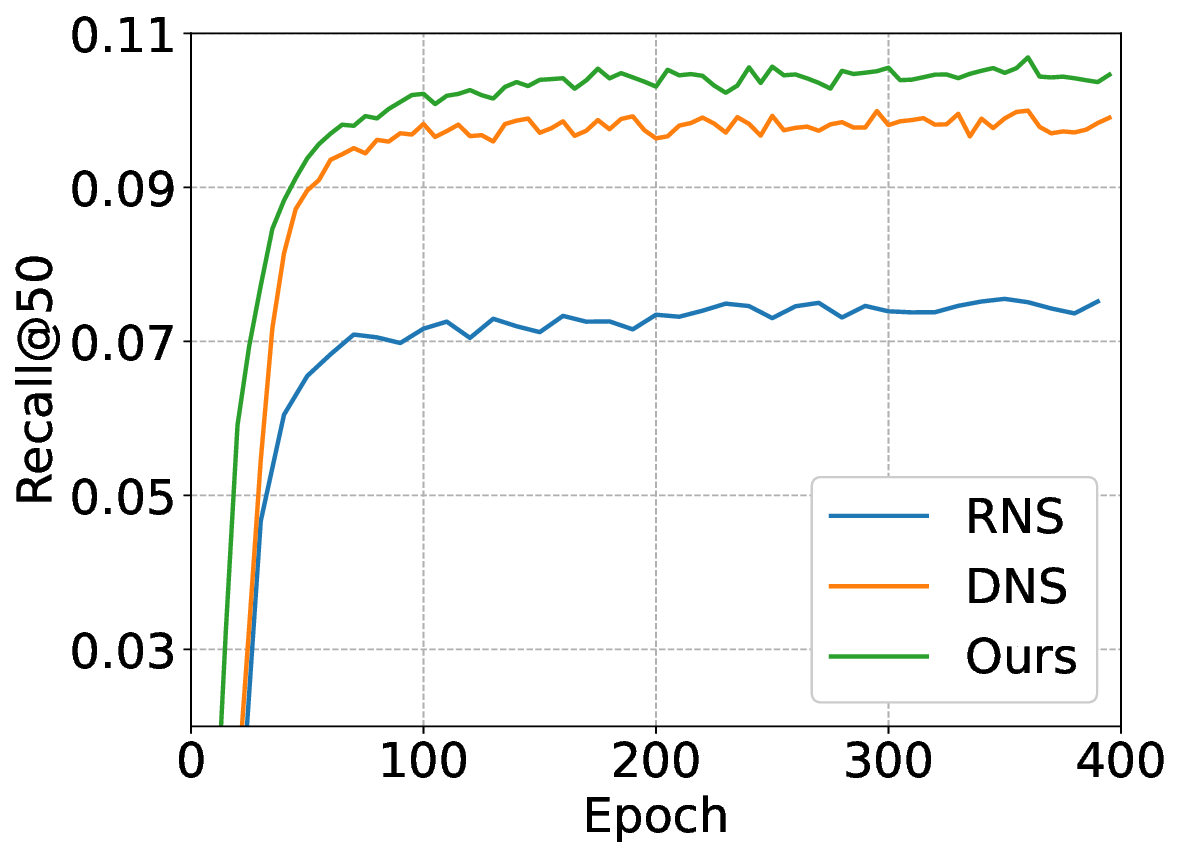}}
\subfigure[MF, Tmall]{
\includegraphics[width=0.55\columnwidth]{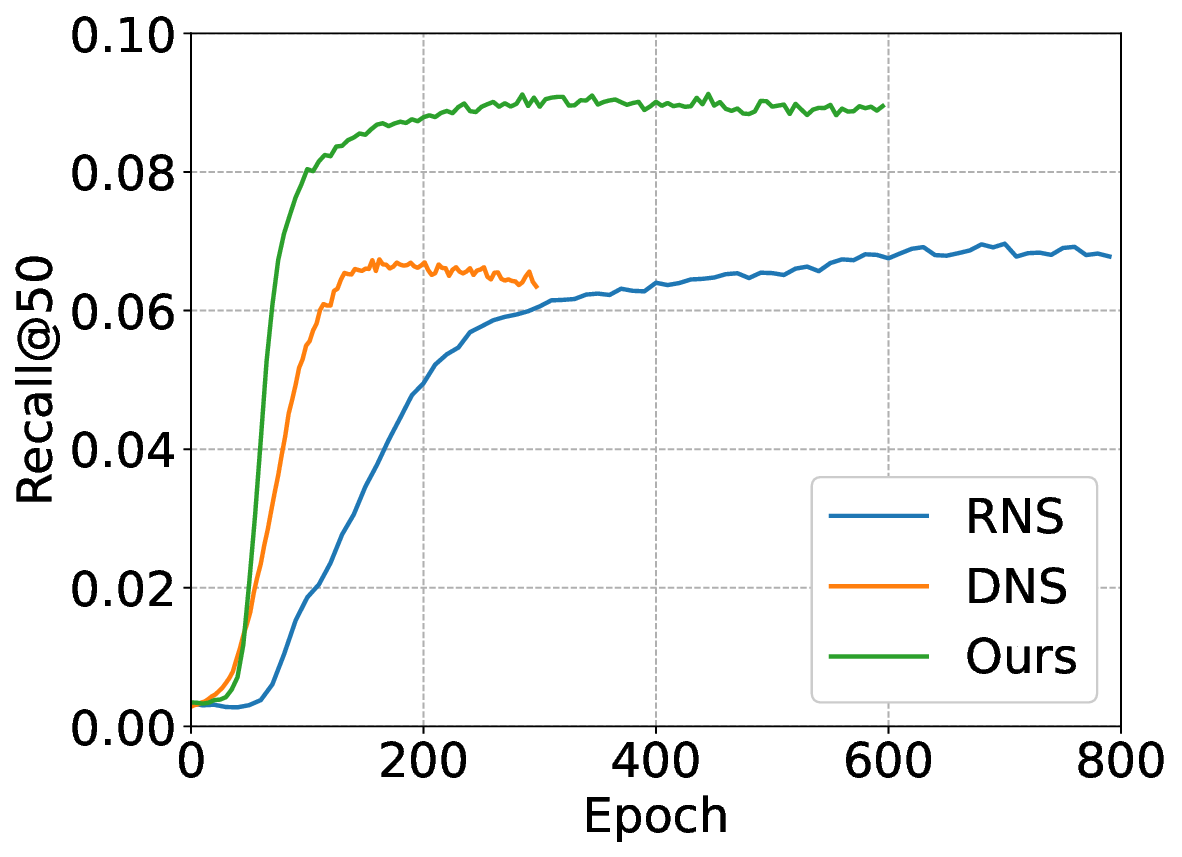}}
\subfigure[MF, Gowalla]{
\includegraphics[width=0.55\columnwidth]{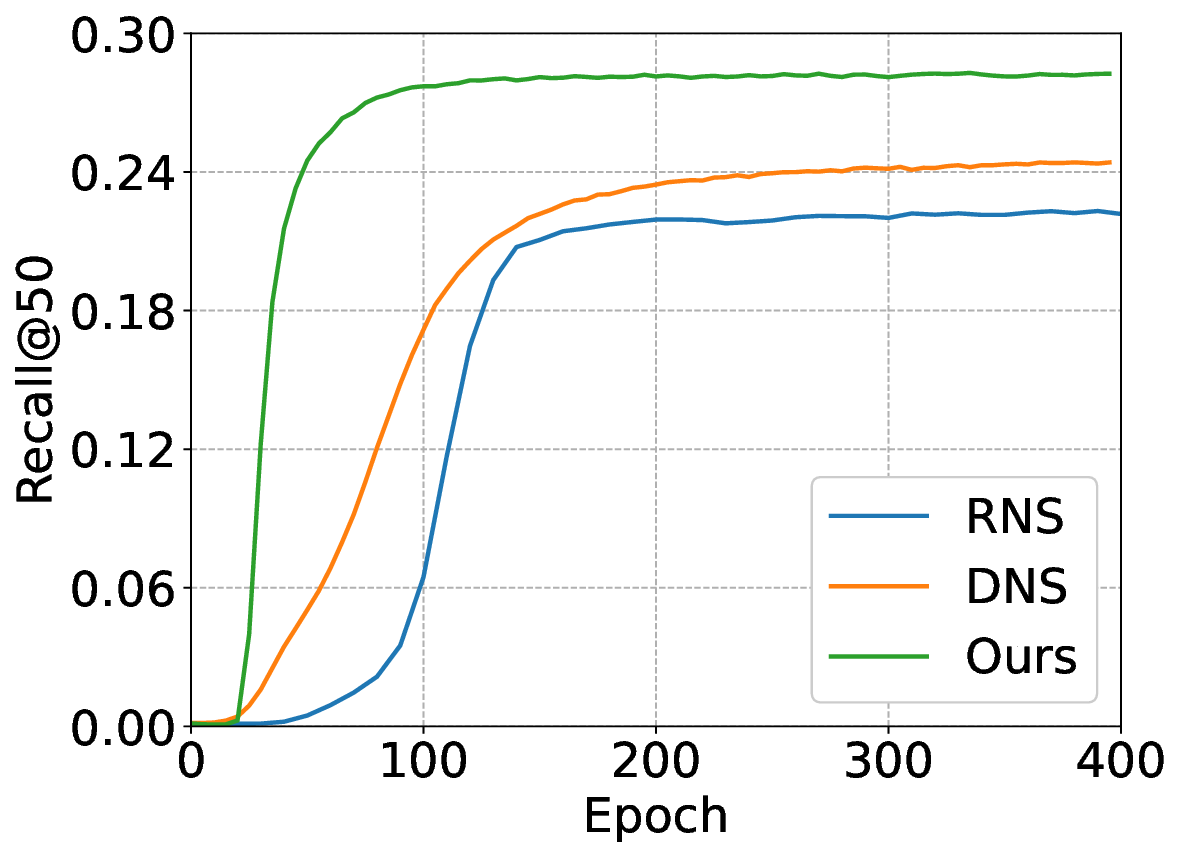}}
\subfigure[LightGCN, Taobao]{
\includegraphics[width=0.55\columnwidth]{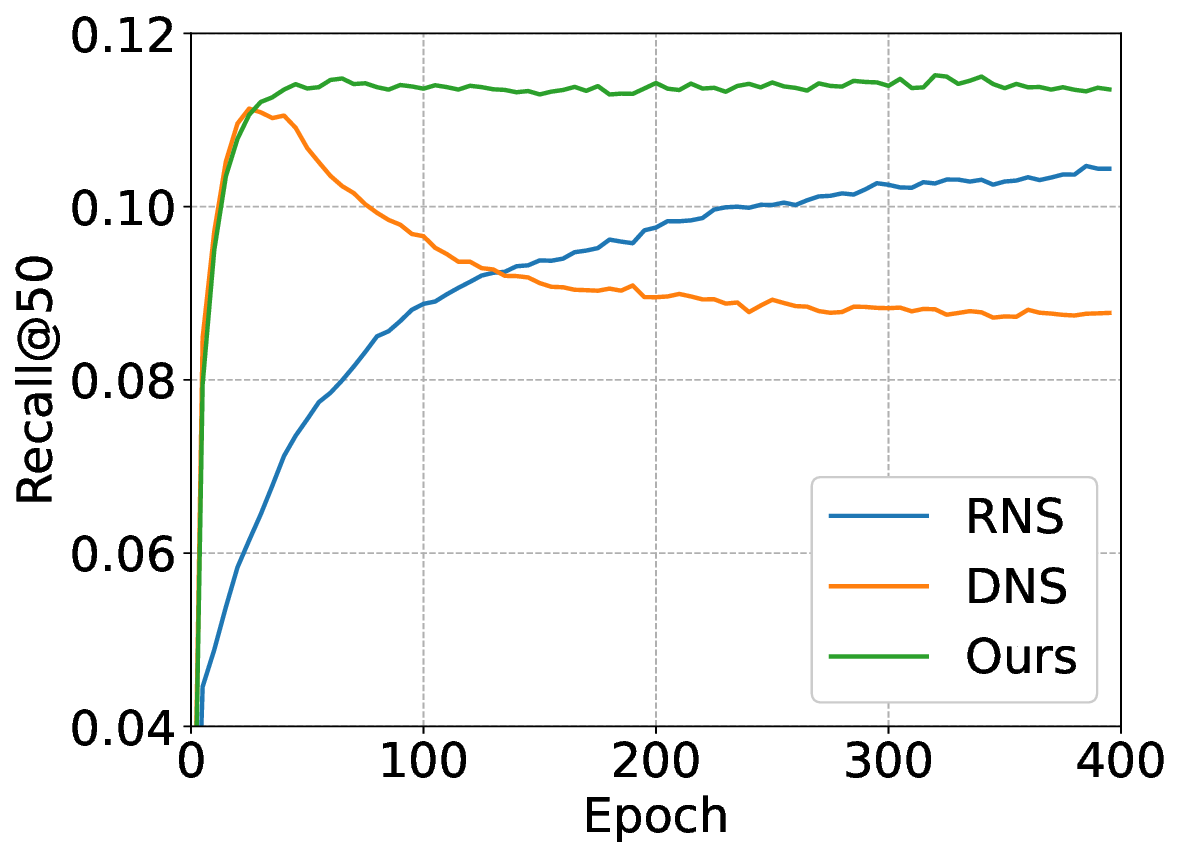}}
\subfigure[LightGCN, Tmall]{
\includegraphics[width=0.55\columnwidth]{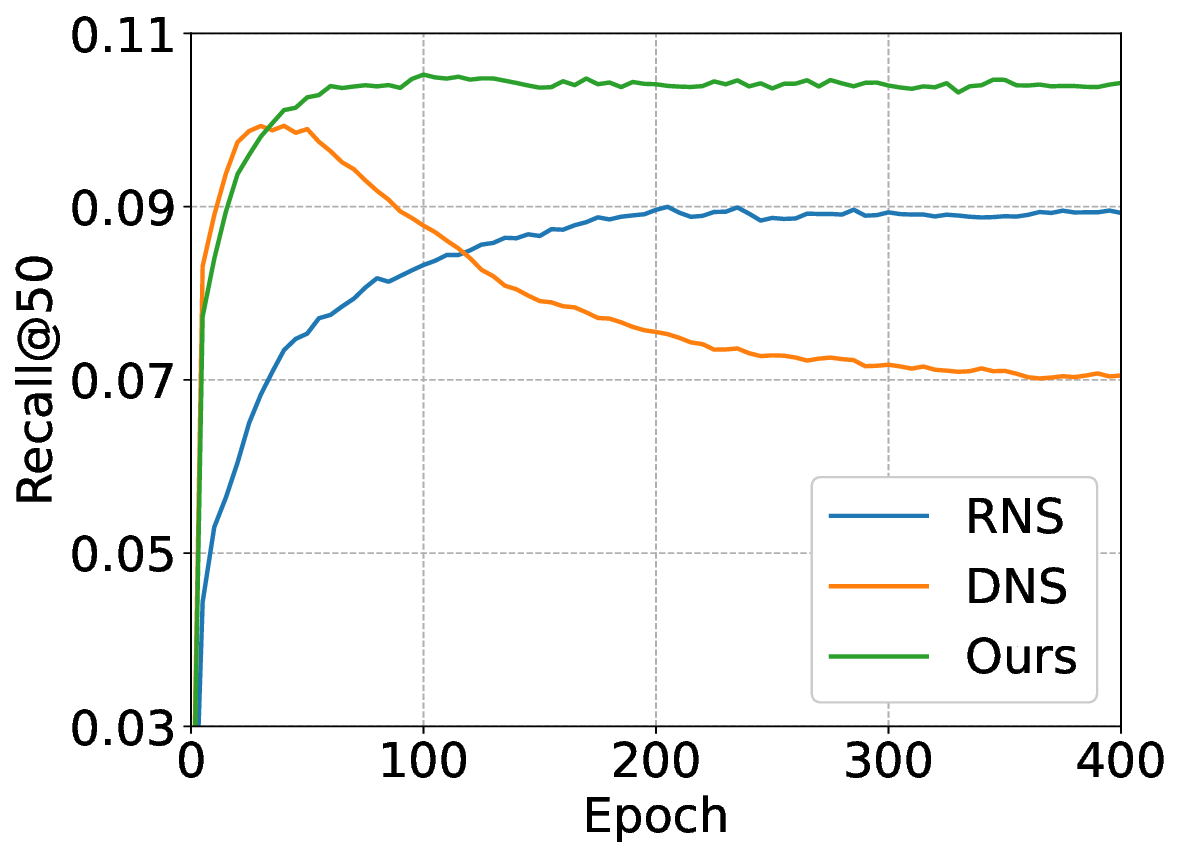}}
\subfigure[LightGCN, Gowalla]{
\includegraphics[width=0.55\columnwidth]{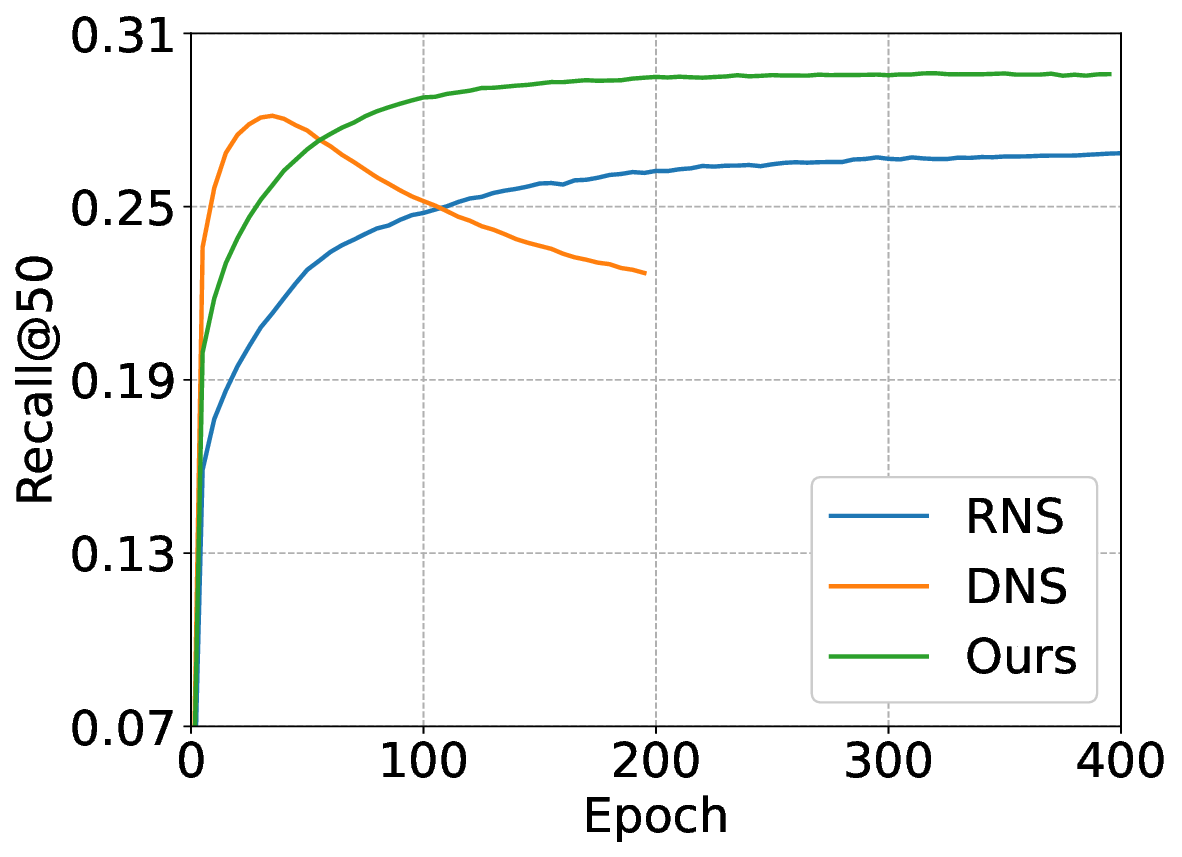}}
\caption{The training curves of RNS, DNS and the proposed method on the test sets of three datasets. (a)(b)(c) With MF as the base model. (d)(e)(f) With LightGCN as the base model.}
\label{fig:rns_dns_ours}
\end{figure*}

To compare different hard mining methods, we conduct experiments on Taobao, Tmall and Gowalla, employing MF or LightGCN as the base recommendation model, respectively. The results, presented in Table~\ref{tab:comparison_mf_lgn}, yield the following observations:
\begin{itemize}
    \item Across both recommendation models, MF and LightGCN, our method consistently outperforms all baselines on the three datasets in terms of Recall@50 and NDCG@50. This highlights the effectiveness of the proposed Hard-BPR loss in a hard negative sampling scenario. Besides, DNS acts as a competitive approach, achieving a top performance in two out of six comparisons among baselines.
    \item The average relative improvement of our method over the best baseline under MF is 7.40\% for NDCG@50, surpassing the improvement under LightGCN, which is 3.07\%. This difference arises because Hard-BPR facilitates MF to learn more fine-grained features, thereby unleashing the predictive potential inherent in this simple model. 
\end{itemize}

Furthermore, we present the training curves of RNS, DNS, and our method under MF and LightGCN, shown in Figure~\ref{fig:rns_dns_ours}. The key findings are outlined as follows:
\begin{itemize}
    \item Compared to RNS, the hard negative sampling method DNS converges (or reaches its maximum value if overfitting occurs) faster and achieves a higher Recall@50. However, the application of DNS in recommender systems may lead to overfitting, especially when LightGCN serves as the underlying model. As studied in previous works~\cite{shi2022soft}, overfitting may result from the inclusion of false negatives during hard negative sampling. In addition, overfitting is more severe under LightGCN compared to MF. This is because the optimal hyperparameter $L_2$ regularization under MF is larger than the one under LightGCN in experiments, thus preventing overfitting to some extent.
    \item By replacing $\sigma(\cdot)$ with $g(\cdot)$, overfitting is effectively eliminated as expected. It verifies that Hard-BPR exhibits the capacity to mitigate the influence of false negatives in hard mining process. Our method is thereby demonstrated as a robust hard negative sampling method in implicit CF.
    \item The training efficiency of our method is comparable to that of DNS, both of which exhibit the lowest time complexity among baselines as shown in Table~\ref{tab:time_complexity}. Notably, when MF acts as the base recommendation model, our method even converges faster than DNS. This suggests our method's efficiency, which facilitates its deployment in realistic scenarios.
\end{itemize}

\subsection{False negative analysis}

\begin{figure*}[h]
\centering
\subfigure[MF, RNS]{
\includegraphics[width=0.55\columnwidth]{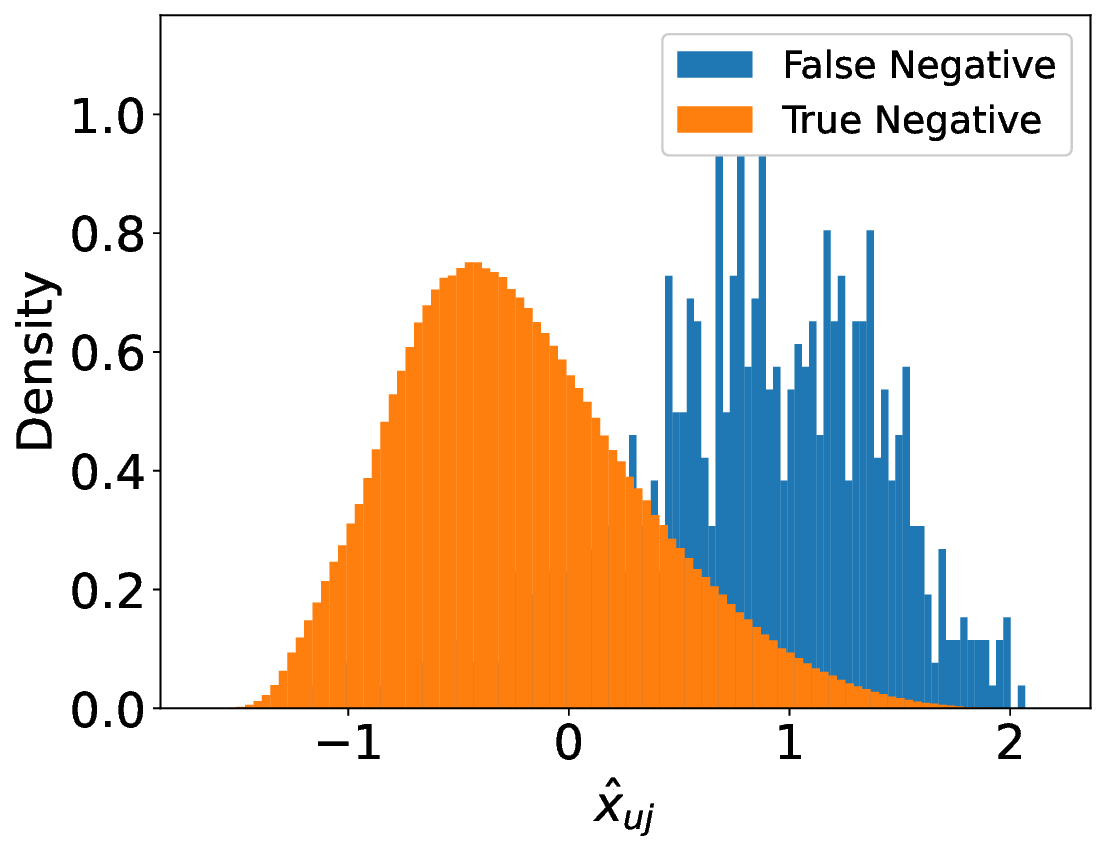}}
\subfigure[MF, DNS]{
\includegraphics[width=0.55\columnwidth]{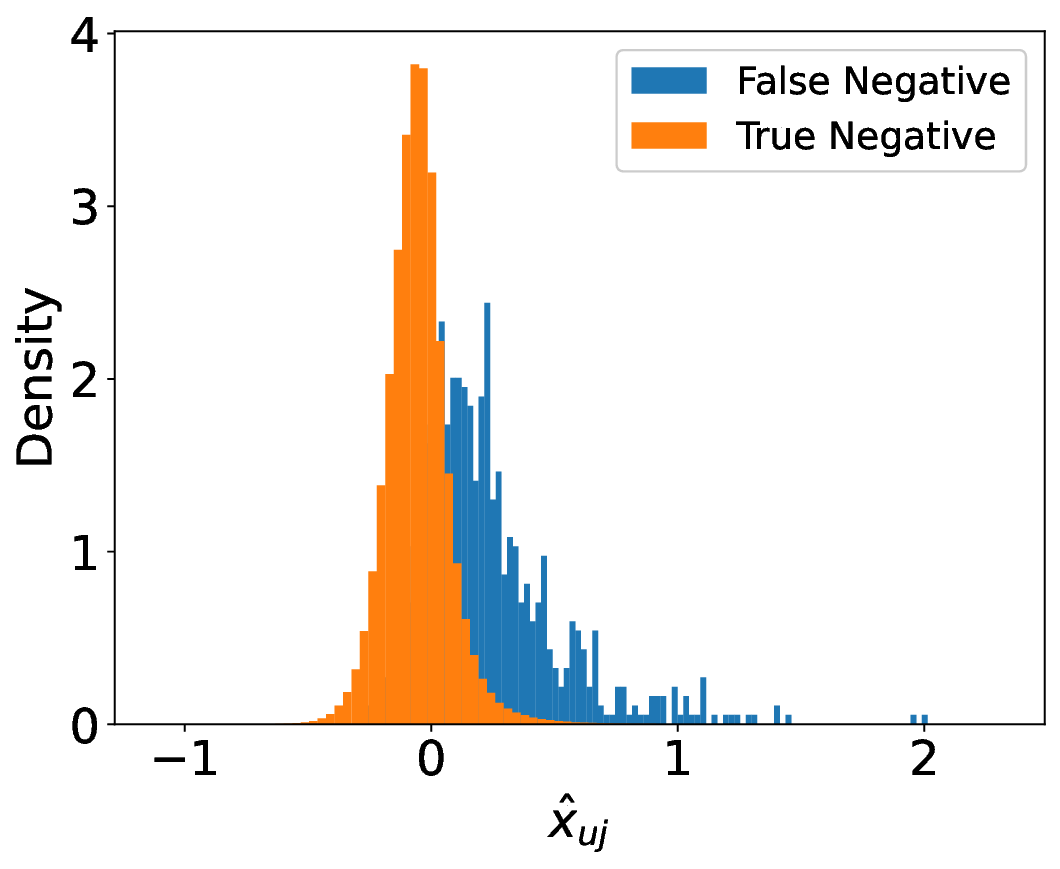}}
\subfigure[MF, our method]{
\includegraphics[width=0.55\columnwidth]{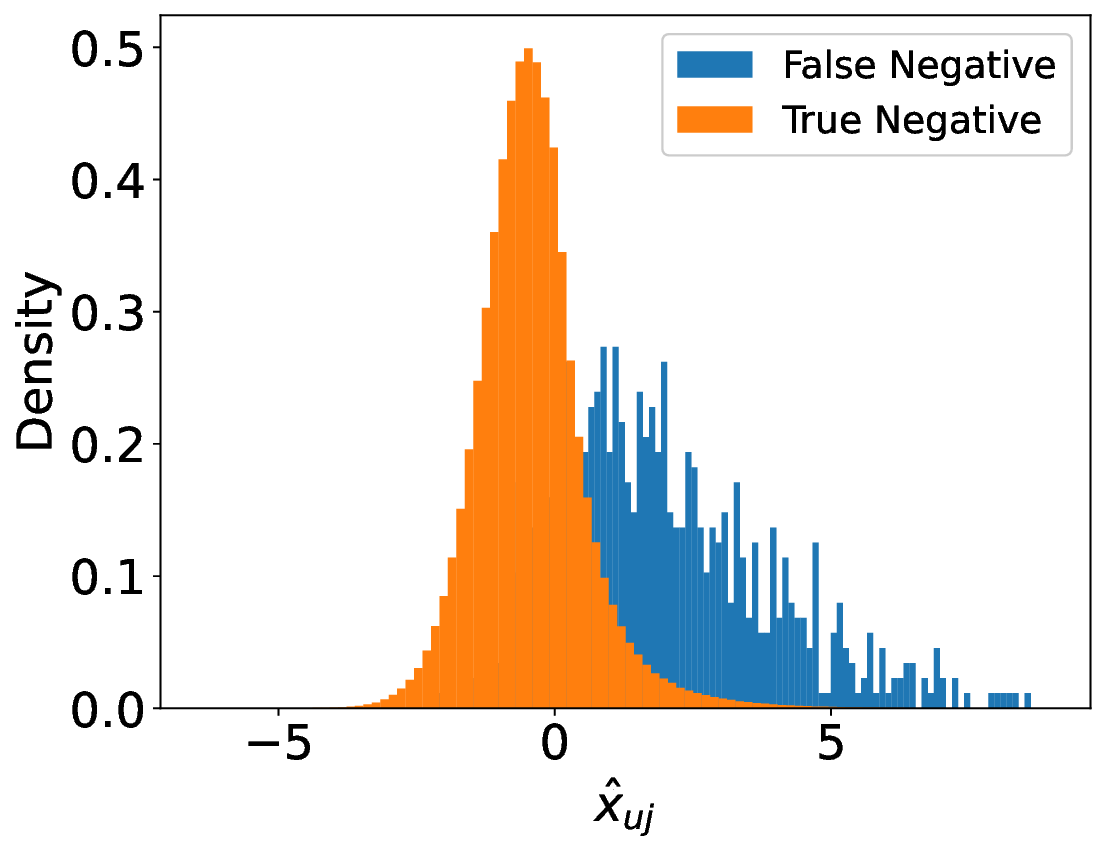}}
\subfigure[LightGCN, RNS]{
\includegraphics[width=0.55\columnwidth]{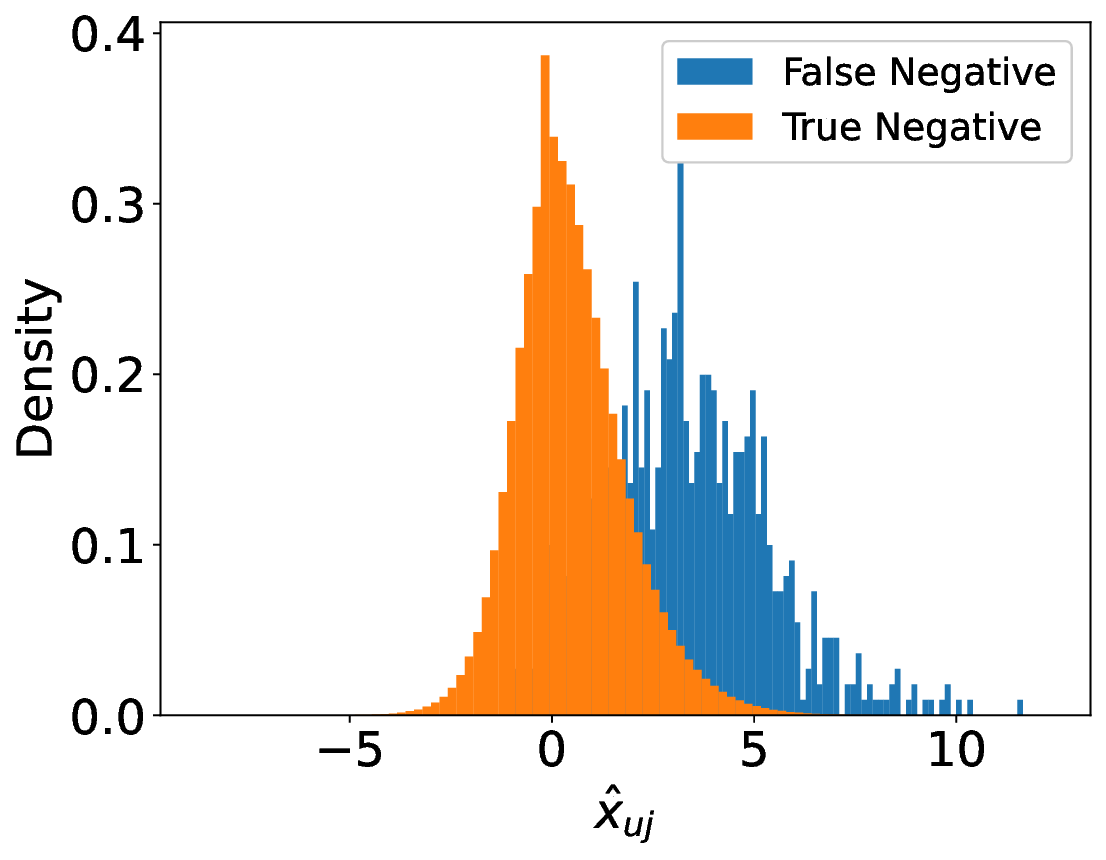}}
\subfigure[LightGCN, DNS]{
\includegraphics[width=0.55\columnwidth]{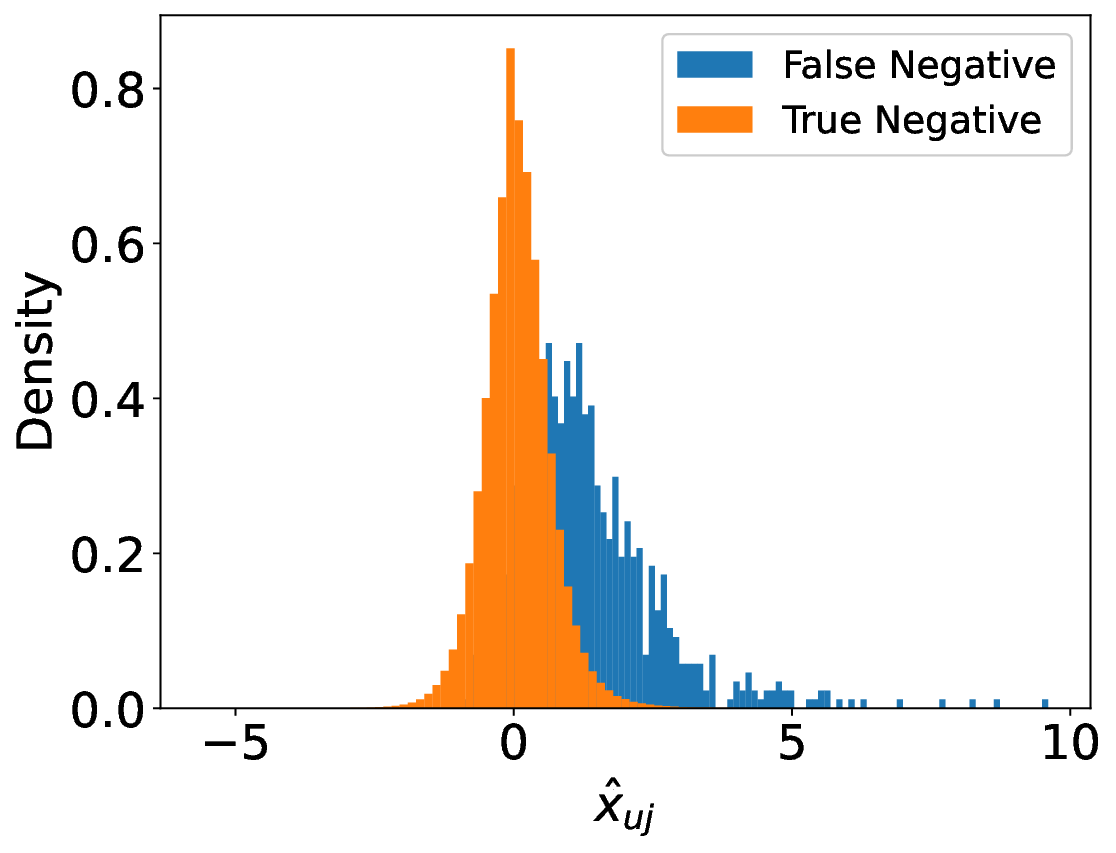}}
\subfigure[LightGCN, our method]{
\includegraphics[width=0.55\columnwidth]{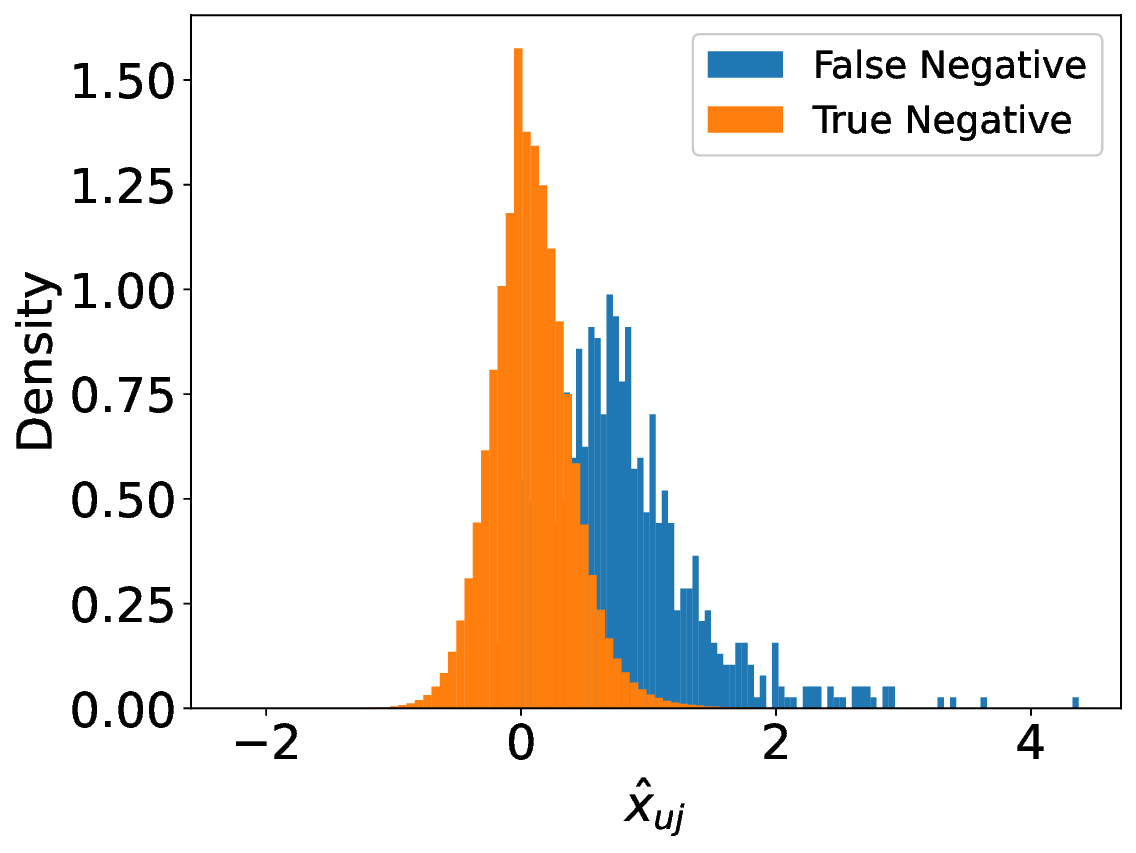}}
\caption{The distributions of predicted true negative scores and predicted false negative scores, employing different base recommendation models and different negative sampling methods (RNS, DNS and the proposed method).}
\label{fig:false_distinguish}
\end{figure*}

False negatives generally indicate a user’s potential preferences. Accurately identifying false negatives is critical for recommending items that the user would have liked or interacted with. In the preceding section, we validate that Hard-BPR effectively prevents the model from being misled by incorrect information resulting from false negatives. In this section, we evaluate the recommender system’s capability to distinguish false negatives from true negatives. 
Experiments on Taobao employ MF and LightGCN as the base models, comparing three negative sampling methods: RNS, DNS and our method. For each user in Taobao, we treat their interacted items in the test set as false negatives in model training process and the remaining non-interacted items as true negatives. Models are saved at the point of achieving the best validation Recall@50. For each user, we compute his (or her) matching scores with true negative instances (yellow) and false negative instances (blue), plotting their distributions in Figure~\ref{fig:false_distinguish}. To clearly manifest the discriminative capability of three negative sampling methods, we apply Gaussian kernel density estimation to estimate the probability distributions of true negative scores and false negative scores. We then calculate the Kullback-Leibler (KL) divergence between these two probability distributions, as presented in Table~\ref{tab:KL}. 

\begin{table}[htp]
  \begin{center}
    \caption{The Kullback-Leibler divergence between the distributions of predicted true negative scores and predicted false negative scores.}
    \label{tab:KL}
    \begin{adjustbox}{width=0.5\textwidth}
        \begin{tabular}{l|ccc|ccc}
        \hline & \multicolumn{3}{c|}{ MF } & \multicolumn{3}{c}{ LightGCN }\\
        & RNS & DNS & Ours & RNS & DNS & Ours \\
        \hline
        \multirow{1}{*}{KL divergence}
        & 1.32216 & 1.35012 & \underline{1.43378} & 1.46923 & 1.35639 & \underline{1.50273} \\
        \hline
        \end{tabular}
    \end{adjustbox}
  \end{center}
\end{table}

We find that the KL divergence under LightGCN averages 1.44278 across RNS, DNS and our method, surpassing that under MF, which is 1.36869. This observation indicates the superior ability of LightGCN to distinguish false negatives compared to MF. With MF as the base model, our method exhibits a 6.20\% relative improvement in KL divergence over DNS, while with LightGCN, our method achieves a relative improvement of 10.79\% over DNS. It verifies the capacity of Hard-BPR in false negative identification during hard mining process.
Notably, when employing LightGCN, the KL divergence between the two distributions is merely 1.35639 under DNS, even smaller than that under RNS (1.46923). This discrepancy may be attributed to the adverse impact of false negatives on model updating with DNS under LightGCN, which not only gives rise to severe overfitting as shown in Figure~\ref{fig:rns_dns_ours}(d)-(f) but also diminishes the model’s discriminative ability for false negatives. 

\subsection{Parameter study}

\begin{figure}[htp]
\centering
\subfigure[The position effect study]{
\label{fig.parameter_study_1}
\includegraphics[width=\columnwidth]{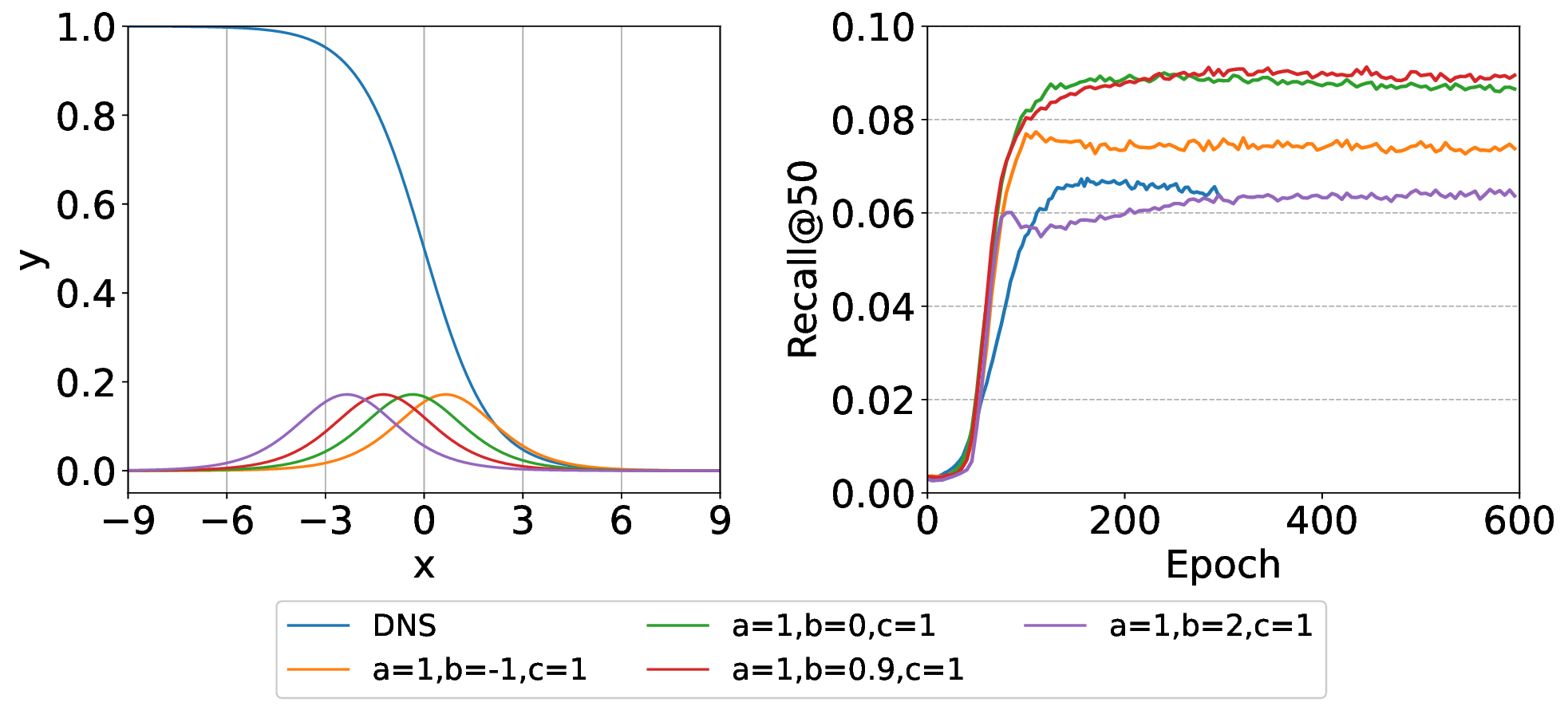}}
\subfigure[The curve shape effect study]{
\label{fig.parameter_study_2}
\includegraphics[width=\columnwidth]{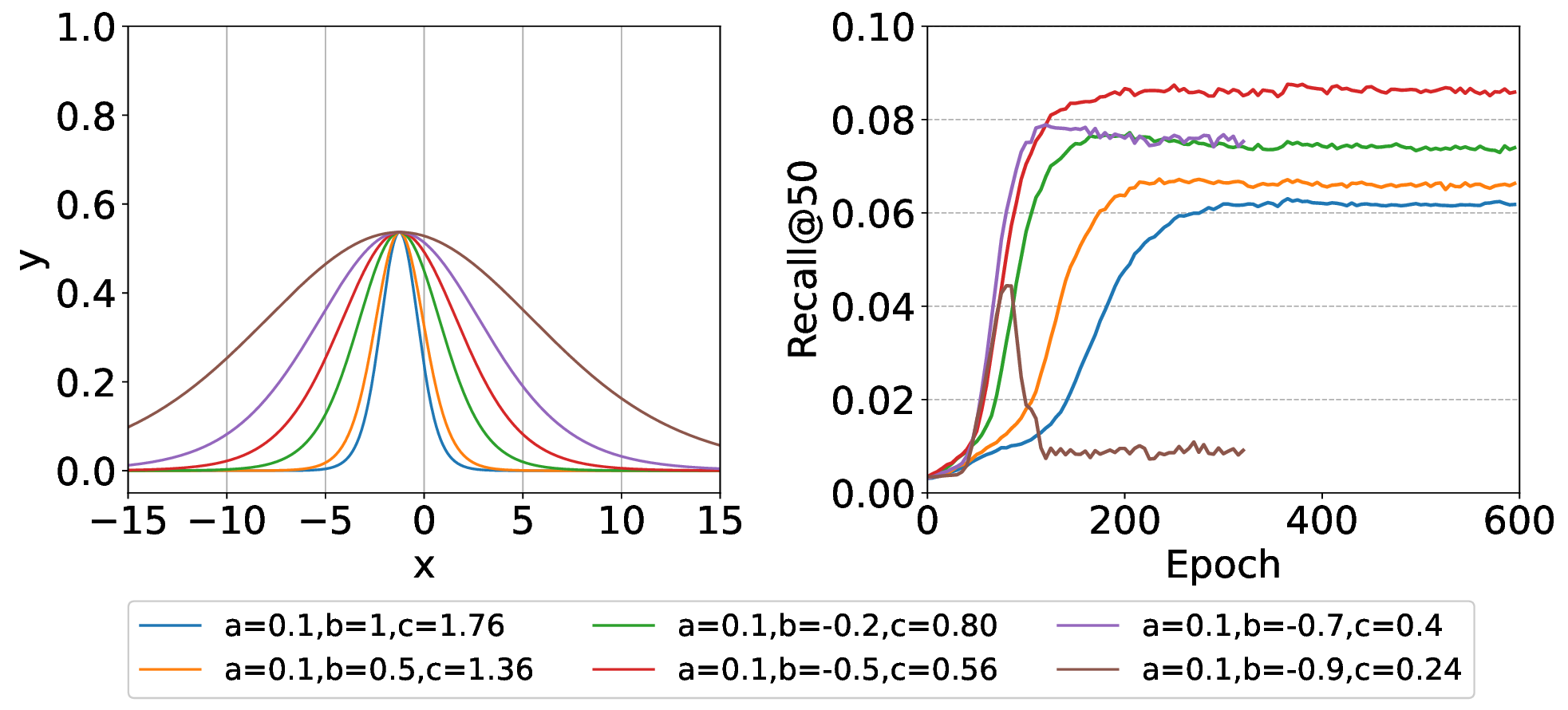}}
\subfigure[Under various $a$ settings]{
\label{fig.parameter_study_3}
\includegraphics[width=\columnwidth]{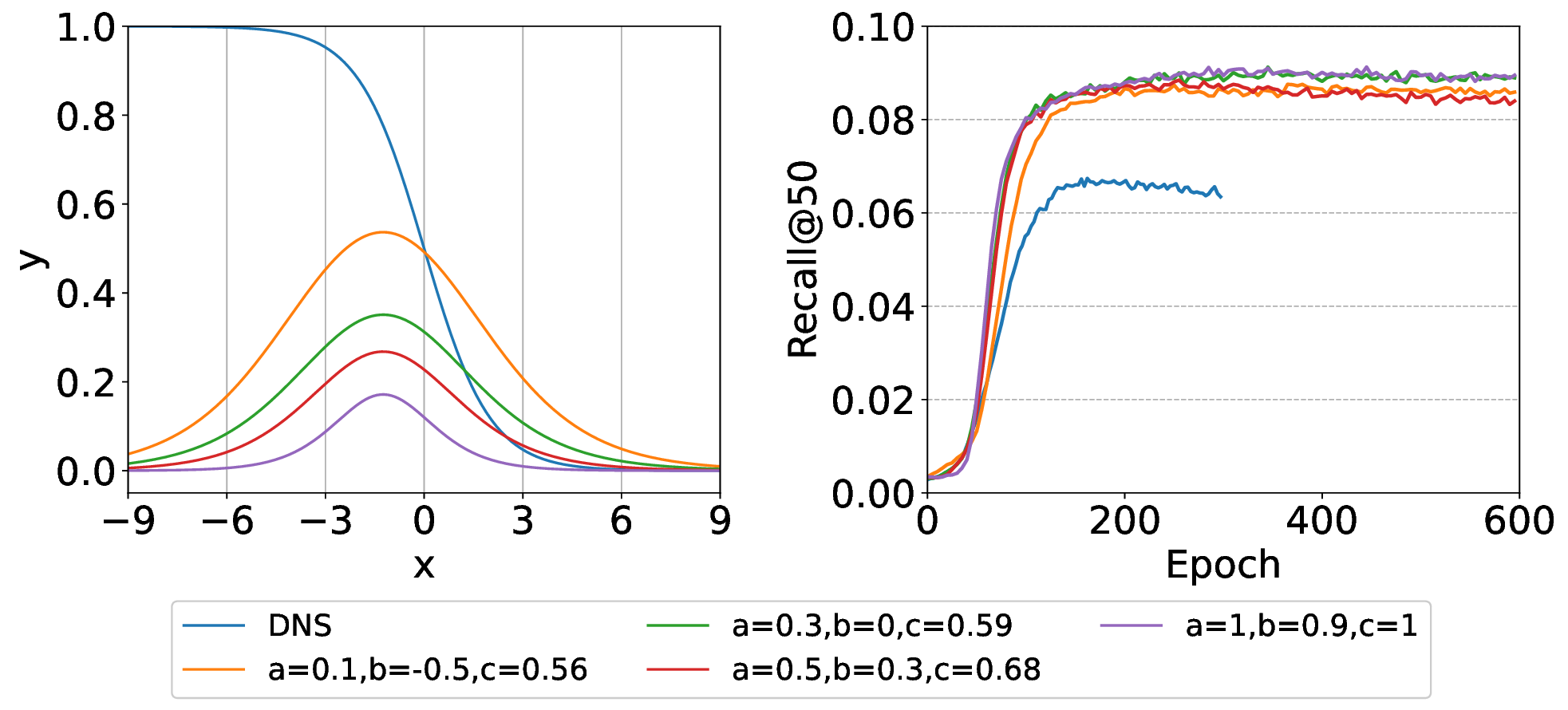}}
\caption{Parameter study of the proposed method. Experiments on Tmall under MF. The left column: curves of $\Delta_g(\cdot)/c$. The right column: training curves of our method. }
\end{figure}


While the implementation of our method alleviates the need for tunning sophisticated sampling process, there remains a necessity to finely adjust coefficients $a, b\text{ and } c$ $(a \in [0,+\infty)$, $b\in (-\infty, +\infty)$, and $c \in (0, +\infty))$ in Hard-BPR loss. In this section, we study the effect of these three coefficients on model performance to provide insights for fine-tuning the model. Experiment results on Tmall under MF are presented in Figure~\ref{fig.parameter_study_1}, Figure~\ref{fig.parameter_study_2} and Figure~\ref{fig.parameter_study_3}. The left column of Figures shows the curves of $\Delta_g(\cdot)/c$ under various settings (for gradient magnitude, dividing by a constant has no influence on model training under Adam).

In Figure~\ref{fig.parameter_study_1}, we set $a\text{ and }c$ as 1 and vary the $b$ value to investigate the effect of $\Delta_g$ curve’s position on model performance. We observe that the model performs optimally when $b$ is 0 or 0.9, while suboptimal performance is observed when the $\Delta_g$ curve is positioned towards the far left or far right. Hence, ensuring the $\Delta_g$ curve concentrated within an appropriate user-item score range is critical for effective model training. In addition, compared to DNS method, our method with $b$ as 0.9 exhibits faster convergence, higher Recall@50 and mitigated overfitting.

Both $b \text{ and } c$ affect the position of $\Delta_g$ curve along x-axis, while $c$ also controls the curve’s shape. In Figure~\ref{fig.parameter_study_2}, we fix the coefficient $a$ to 0.1, and vary the curve from a sharp to a broader shape by reducing the coefficient $c$. Meantime, we correspondingly adjust the $b$ to maintain the maximum point $x_{\text{max}}$ of curves as the same. We observe that a sharp $\Delta_g$ curve results in a bad performance in model effectiveness and efficiency, because it places excessive importance on a small range of $\hat{x}_{uij}$ around $x_{\text{max}}$ for model learning. It leads to information loss from other hard negatives. Conversely, when the curve becomes boarder under a smaller $c$, more hard negatives can participate in parameter updating, thereby enhancing learning speed and model performance. Notably, excessive extension of the curve tails can also lead to performance deterioration. For example, at a $c$ value of 0.4 or 0.24, the training curve initially peaks and then declines. 

Figure~\ref{fig.parameter_study_3} demonstrates that under various $a$ settings, similar $\Delta_g$ curves can be obtained by adjusting $b\text{ and }c$, leading to comparable performance. The coefficient $a$ is critical to make the $\Delta_g$ curve bell-shaped, but it is not hard to tune. In practice, we can manually set $a$ to a small value such as 1, while fine-tuning is primarily required for $b\text{ and }c$.

\section{Related work}
Negative sampling approaches in implicit collaborative filtering are developed into several branches in recent years. The simplest way is to follow a static distribution such as uniform distribution~\cite{rendle2012bpr} or popularity-based distribution~\cite{chen2017sampling}, however, which may encounter challenges such as gradient vanishing or fail to adapt to dynamic changes in predicted user preferences. To address these limitations and enhance the quality of sampled negatives, various hard negative sampling strategies, including GAN-based samplers (e.g., IRGAN~\cite{wang2017irgan}, AdvIR~\cite{park2019adversarial}), DNS and its extensions, and softmax-based sampling approaches, have been proposed. GAN-based samplers leverage GANs to iteratively generate informative negative instances. DNS~\cite{zhang2013optimizing}, a simple yet effective framework, involves random sampling of negative candidates for each user and subsequent selection of the highest-scored candidate for model updating. Extensions of DNS, such as MixGCF~\cite{huang2021mixgcf}, SRNS~\cite{ding2020simplify}, GDNS~\cite{zhu2022gain}, and DNS($M,N$)~\cite{shi2023theories}, have further contributed to the evolution of hard negative sampling strategies. Softmax-based samplers, such as AdaSIR~\cite{chen2022learning} or Softmax-v($\rho,N$)~\cite{shi2023theories}, assign resampling weights to each negative candidate. This strategy aims to align the negative sampling probability with the ideal softmax distribution, where greater weights are typically allocated to negative samples that are more informative. Moreover, \citet{shi2023theories} proves that the BPR loss combined with DNS is an exact estimator of OPAUC, and they extend DNS to DNS($M,N$) to suit various Top-K metrics. In our study, without applying important resampling, we directly modify the loss function to ensure the contribution of hard negatives on parameter updating.

For hard negative sampling, the inadvertent inclusion of false negatives poses a challenge on effective model learning. \citet{cai2022hard} formulates the false negative problem as learning from labeled data with inherent bias, and adopts a Coupled Estimation Technique to correct the bias. To avoid false negatives, SRNS develops a scored-based and variance-based sampling strategy, and GDNS proposes a gain-aware sampler to distinguish true hard negatives. However, both SRNS and GDNS center their efforts on crafting sophisticated sampling progress to avoid the inclusion of false negatives, lacking a desired level of elegance. The investigation of overfitting related to false negatives in the study ~\cite{shi2022soft} provides valuable insights into our research.

\section{Conclusion}
This paper aims to address the false negative problem of hard negative sampling in implicit collaborative filtering. Inspired by the equivalence theory, our approach diverges from focusing on the sampling process design. Instead, we put efforts on loss function crafting, proposing an enhanced BPR objective, Hard-BPR, specifically tailored for dynamic hard negative sampling. Employing DNS under Hard-BPR is simple and efficient enough for real-world deployments. This approach not only demonstrates resilience to false negatives during hard sampling, but also exhibits a notable advantage in terms of time complexity. Moreover, equipping DNS with Hard-BPR serves as an exact estimator of OPAUC. Extensive experiments are conducted to assess the proposed method’s robustness and effectiveness, and the parameter study is conducted to provide guidance for optimal tunning. 

\appendix
\section{appendix}
\subsection{Derication of $\Delta_{g}(\cdot)$ properties.} 
The fist derivative of $\Delta_g(x)$ is calculated as:
\begin{equation}
\begin{aligned}
\Delta_g^{’}(x) &= \frac{c^2 \left(1-\frac{1}{e^{-b-c x}+1}\right) e^{-b-c x}}{\left(e^{-b-c x}+1\right)^2 \left(a+\frac{1}{e^{-b-c x}+1}\right)}-\frac{c^2 e^{-b-c x}}{\left(e^{-b-c x}+1\right)^3 \left(a+\frac{1}{e^{-b-c x}+1}\right)}\\& -\frac{c^2 e^{-b-c x} \left(1-\frac{1}{e^{-b-c x}+1}\right)}{\left(e^{-b-c x}+1\right)^3 \left(a+\frac{1}{e^{-b-c x}+1}\right)^2} \\
&= -\frac{c^2 e^{b+c x} \left(a \left(e^{2 (b+c x)}-1\right)+e^{2 (b+c x)}\right)}{\left(e^{b+c x}+1\right)^2 \left(a e^{b+c x}+a+e^{b+c x}\right)^2}
\end{aligned}
\end{equation}
By solving $\Delta_g^{’}(x)=0$, the critical point of $\Delta_g(x)$ is calculated as $x_{\text{c}} = \left(-b+\ln(\frac{\sqrt{a}}{\sqrt{1+a}})\right)/c$. Since the graph of $\Delta_g(x)=0$ $(a>0)$ is unimodal, $x_{\text{c}}$ is the maximum point, termed as $x_{\text{max}}$.

To prove the symmetry about the maximum point $x_{\text{max}}$:
\begin{equation}
\begin{aligned}
\Delta_g(x+ x_{\text{max}}) &= \frac{c \left(1-\frac{1}{e^{-c \left(\frac{\log \left(\frac{\sqrt{a}}{\sqrt{a+1}}\right)-b}{c}+x\right)-b}+1}\right)}{\left(e^{-c \left(\frac{\log \left(\frac{\sqrt{a}}{\sqrt{a+1}}\right)-b}{c}+x\right)-b}+1\right) \left(\frac{1}{e^{-c \left(\frac{\log \left(\frac{\sqrt{a}}{\sqrt{a+1}}\right)-b}{c}+x\right)-b}+1}+a\right)}\\
&= \frac{\sqrt{\frac{a}{a+1}} c e^{c x}}{\left(\sqrt{\frac{a}{a+1}} e^{c x}+1\right) \left(\sqrt{\frac{a}{a+1}} a e^{c x}+\sqrt{\frac{a}{a+1}} e^{c x}+a\right)}
\end{aligned}
\end{equation}

\begin{equation}
\begin{aligned}
\Delta_g(-x+x_{\text{max}}) &= \frac{c \left(1-\frac{1}{e^{-c \left(\frac{\log \left(\frac{\sqrt{a}}{\sqrt{a+1}}\right)-b}{c}-x\right)-b}+1}\right)}{\left(e^{-c \left(\frac{\log \left(\frac{\sqrt{a}}{\sqrt{a+1}}\right)-b}{c}-x\right)-b}+1\right) \left(\frac{1}{e^{-c \left(\frac{\log \left(\frac{\sqrt{a}}{\sqrt{a+1}}\right)-b}{c}-x\right)-b}+1}+a\right)}\\
&= \frac{\sqrt{\frac{a}{a+1}} c e^{c x}}{\left(\sqrt{\frac{a}{a+1}}+e^{c x}\right) \left(a \left(\sqrt{\frac{a}{a+1}}+e^{c x}\right)+\sqrt{\frac{a}{a+1}}\right)}
\end{aligned}
\end{equation}
Hence, we have $\Delta_g(x+x_{\text{max}})=\Delta_g(-x+x_{\text{max}})$.


\clearpage

\bibliographystyle{ACM-Reference-Format}
\bibliography{DNS-BPR}

\end{document}